\documentclass[11pt,a4paper]{article}
\pdfoutput=1
\usepackage{jheppub}
\usepackage{lscape}
\usepackage{array}
\usepackage{hyperref}
\usepackage{amsfonts}
\usepackage{amssymb}
\usepackage{bbm} 
\usepackage{graphicx}
\usepackage{caption}
\usepackage{subcaption}

\usepackage[normalem]{ulem}

\newcommand{\X}{X}

\newcommand{\diff}{\text{d}}
\renewcommand{\=}{\,= \,}

\renewcommand{\a}{\alpha}

\newcommand{\liMN}{\underset{\tau\to-\frac{n}{m}}{\simeq}}

\renewcommand{\i}{{\rm i}}

\newcommand{\e}{{\bf e}}

\newcommand\bi{\begin{itemize}}
\newcommand\ei{\end{itemize}}

\newcommand\bspl{\begin{split}}
\newcommand\espl{\end{split}}

\newcommand{\susyL}[1]{\underset{\text{SUSY Locus}}{\longrightarrow}}

\newcommand{\Ethi}{(E_a)_{\theta}}

\newcommand{\be}{\begin{equation}}
\newcommand{\ee}{\end{equation}}
\newcommand{\bea}{\begin{eqnarray}}
\newcommand{\eea}{\end{eqnarray}}

\renewcommand{\=}{\,= \,}

\renewcommand{\a}{\alpha}

\renewcommand{\i}{{\rm i}}

\renewcommand{\e}{{\bf e}}

\newcommand{\limNorth}{\underset{g\,\to\, 0^+}{\longrightarrow}}
\newcommand{\limSouth}{\underset{g\,\to\, \frac{\pi}{2}-0^+}{\longrightarrow}}

\usepackage{amsmath}

\begin{document}

\title{Quantum Phases of~$4d$~$SU(N)$~$\mathcal{N}=4$ SYM}

\author{Alejandro Cabo-Bizet}
\emailAdd{alejandro.cabo\_bizet@kcl.ac.uk}
\affiliation{Department of Mathematics, King's College London,\\
The Strand, London WC2R 2LS, U.K.}

\abstract{It is argued that~$4d$~$SU(N)$~$\mathcal{N}=4$ SYM has an accumulation line of zero-temperature topologically ordered phases. Each of these phases corresponds to~$N$ bound states charged under electromagnetic~$\mathbb{Z}^{(1)}_N$ one-form symmetries. Each of the~$N$ bound states is made of two Dyonic flux components each of them extended over a two dimensional surface. They are localized at the fixed loci of a rotational action, and are argued to correspond to conformal blocks (or primaries) of an~$SU(N)_1$ WZNW model on a two-torus. 

}

\date{\today}

\maketitle

\section{Introduction}
\label{sec1}

Understanding the zero-temperature properties of a physical system is a fundamental problem in theoretical physics. Such a limit is expected to be far from trivial for systems with a large ground-state degeneracy. In the context of AdS/CFT duality~\cite{Maldacena:1997re} varying the temperature of a conformal field theory (CFT) corresponds to deforming the boundary conditions that fix the dual AdS string/gravitational observable. Sometimes the latter observable can be obtained from an action functional evaluated at a classical gravitational configuration. When this dual configuration is a black hole, the temperature of the CFT is naturally identified with the Bekenstein-Hawking temperature~\cite{Witten:1998zw}. Other chemical potentials are identified as well, on both sides of the duality, by matching the background geometry in which the CFT is quantized upon, against the boundary conditions fixed by the dual~AdS geometry~\cite{Benini:2015eyy,Cabo-Bizet:2018ehj}.

In the context of~AdS$_5$/CFT$_4$ a zero-temperature limit reaching extremal and supersymmetric (BPS) black hole solutions~\cite{Gutowski:2004ez} within a larger family of non-extremal and non-supersymmetric ones~\cite{Cvetic:2004hs,Cvetic:2004ny}, has recently gotten attention~\cite{Cabo-Bizet:2018ehj}\cite{Cassani:2019mms,Larsen:2019oll,Kantor:2019lfo}. On the CFT side of the duality, it has been shown~\cite{Choi:2018hmj,Benini:2018ywd,Cabo-Bizet:2018ehj}\cite{Hosseini:2017mds,Honda:2019cio, ArabiArdehali:2019tdm,Kim:2019yrz, Cabo-Bizet:2019osg}\cite{Murthy:2020rbd,Agarwal:2020zwm} that such a limit reduces the statistical properties of the strongly coupled system to that of a simpler subsystem of states preserving the same supercharges as the dual BPS black hole, more precisely of states in the cohomology of such supercharges. In these limits, the thermal partition function reduces to a refinement of a Witten index, known as the superconformal index~\cite{Romelsberger:2005eg,Kinney:2005ej}\cite{Dolan:2008qi}.

In the canonical example of AdS$_5$/CFT$_4$ duality, where the CFT$_4$ is 4d~$SU(N)$~$\mathcal{N}=4$ SYM, the superconformal index is a function of various chemical potentials, among which one finds angular velocities and other potentials dual to R-charges. A simplified version of such an index depends on a single combination of angular velocities and~R-symmetry chemical potentials that we will denote as~$\tau\=\tau_1\,+\,\i\tau_2\,$, with~$\tau_1$ and~$\tau_2$ being its real and imaginary parts. For convergence reasons, the parameter~$\tau$ is assumed to span the upper half complex plane. 

Using the index as a tool, a landscape of exotic phases has been recently identified~\cite{Cabo-Bizet:2019eaf} in certain limits~\cite{Cabo-Bizet:2019eaf,Cabo-Bizet:2020ewf} \cite{Goldstein:2020yvj,Jejjala:2021hlt,ArabiArdehali:2021nsx}. These phases accumulate along the real axis at~$\tau_2\=0\,$, more precisely when~$\tau$ approaches generic rational points within appropriate angular sections. The corresponding limits are called generalized Cardy or Cardy-like limits. They correspond to a large charge/spin semiclassical expansion (independent of the large-$N$ expansion). As shown in~\cite{Cabo-Bizet:2019eaf}, at large~$N$, a Cardy-like limit of the index corresponds to a large R-charge ($Q$) and large angular momentum ($J$) semiclassical expansion of the index, for which the ratio between large vacuum expectation values~$\frac{Q^3}{J^2}$ remains finite.~The same happens at finite $N$~\cite{Ardehali:2021irq,PaperBulk} (as follows from the analysis of Section~\ref{sec:CardyLimit} below). 

Although these results have been found using the superconformal index, in virtue of what was explained above, these phases also emerge in a zero-temperature limit of the physical partition function, at generic gauge coupling; more precisely, in a double limit. The first and dominating one is the zero-temperature limit dual to the gravitational BPS limit above recalled, and the second and subleading one is the Cardy-like limit. A physical or geometrical understanding of the properties of these phases has been missing so far. The aim of this note is to contribute to filling such a gap.

We will argue that these phases are topologically ordered and that the effective theory of vacua is (up to dualities) the gauged~$SU(N)_1$ WZNW model on~$T_2$. These vacua will be shown to correspond to a set of bound states of an operator~$\mathcal{B}$ supported over two disconnected~$T_2$ punctures. Four-dimensional gauge invariance is shown to imply the existence of edge modes at the worldvolume of each of the two-component surface operators. These edge modes organize in representations of an affine~$SU(N)_1$ Kac Moody algebra. Previous results and consistency arguments coming from the representation theory of quantum groups constrain the interactions between the worldvolume theories at each of the two surface operators to be such that the effective theory of~$\mathcal{B}$ becomes topological in the Cardy-like limit. The remaining partition function then is essentially reduced to counting the number of vacua in which the bound field~$\mathcal{B}$ can condensate, which, e.g. in the limit~$\tau\to0$ is~$N$. The bound-state perspective also exists away from the Cardy-like limits, but in that case, one may expect its effective theory to have less constrained dynamics.

The condensates of~$\mathcal{B}$, and thus the~$(m,n)$ phases, are shown to carry fractional electromagnetic flux, with electric flux proportional to~$m$ and magnetic flux proportional to~$n$. They are also shown to be characterized specific electromagnetic~$\mathbb{Z}^{(1)}_N$ one-form charges.

Let us summarize the content of the paper. In section~\ref{TopOrder} we explain what we mean by emergent topological order in Cardy-like limits and set up the stage for the following sections. In section~\ref{sec2} we revisit relevant backup material which has been already developed in the literature. After giving a brief introduction to the concept of one-form symmetry, section~\ref{sec3} proceeds to explain why and how the~$(m,n)$ phases, i.e. the states on which the operator~$\mathcal{B}$ condensates in the Cardy-like limit $\tau\to-\frac{n}{m}$ carry both, electromagnetic flux and electromagnetic~$\mathbb{Z}_N^{(1)}$ one-form charge, the latter being the order parameter of the corresponding phase. To complete the analysis presented in section~\ref{TopOrder}\,, in section~\ref{sec:4}, we show how four-dimensional gauge transformations induce a global~$SU(N)_1$ Kac-Moody action on the phase space of flat connections at a punctured 2-torus, and bootstrap the form of the effective action in the presence of the operator~$\mathcal{B}$ in Cardy-like limit (We do so by using classical results in the literature). In section~\ref{sec:5} we conclude with a summary and some questions for the future.

\section{The origin of topological order}\label{TopOrder}

This initial section explains what it is meant by the emergence of topological order in~$SU(N)$ $\mathcal{N}=4$ SYM. ~\footnote{The driving idea of the discussion presented in this section is closely related to the one used, for instance, in~\cite{Gerasimov:2006zt}, and~\cite{Nekrasov:2009uh,Nekrasov:2009rc}, to infer the relation between supersymmetric gauge theories, quantum groups, and Bethe ansatz construction.} 

For the specific purposes of this manuscript we find convenient to focus the presentation on~$4d$~$\mathcal{N}=4$ SYM. But our conclusions can be applied to more general examples, as the key assumption we rely on is (at least in some limit): 
\begin{itemize}
\item~that the partition function and correlation functions of the corresponding system reduce to integrals over the moduli space of flat connections. 
\end{itemize}
Although the focus of our study, our argument covers other cases such as, for instance, certain limits of non-supersymmetric theories like four-dimensional pure~$SU(N)$ Yang-Mills theory.~\footnote{A case that has been recently studied in~\cite{Antinucci:2022eat}.}

\subsection{The main idea}\label{sec:MainIdea}

The superconformal index of four-dimensional~$\mathcal{N}=4$ SYM can be understood as a path integral of the form
\be\label{PathIntegral}
\mathcal{I}\,=\, \int [DA_\mu][\ldots] \, e^{- \mathcal{Q}V}
\ee
where~$\mathcal{Q}$ is one of the supercharges of~$PSU(2,2|4)$.~The selection of~$\mathcal{Q}$ corresponds to the selection of a superconformal Killing spinor on~$S_1\times S^{\tau}_3$ a twisted version of~$S_1\times S_{3}$ with metric
\be\label{background_metric00Intro}
\diff s^2 \=  \diff t_E^2 +  \diff\theta^2 + \sin^2 \theta\, \bigl(\diff \phi_1+\frac{2\pi\tau}{\beta} \diff t_E\bigr)^2 + \cos^2 \theta \,\bigl(\diff \phi_2 +\frac{2\pi\tau}{\beta} \diff  t_E\bigr)^2 \,,
\ee
that we will comeback to introduce later on around equation~\eqref{background_metric0}.~$\beta$ is the period of the Euclidean time~$t_E\,$, and~$\theta \,\in\, [0, \frac{\pi}{2}]$ and~$\phi_{1,2}\sim \phi_{1,2}+2\pi $.

The~$V$ in~\eqref{PathIntegral} is a $\mathcal{Q}$-odd functional of fields defined in such a way that the part of~$\mathcal{Q}V$ that only includes bosons is semi-positive definite along the contour of integration that defines the path integral.
This integral can be exactly solved~\cite{Nawata:2011un}\cite{Cabo-Bizet:2018ehj} by cohomological path-integration methods~\cite{Witten:1990bs}, a method that has been further developed in the last two decades and it is nowadays known as supersymmetric localization~\cite{Pestun:2007rz}. The localization method reduces the original path integral to an integral over flat connections
\be\label{LocalizationPI}
\int [D A_\mu][\ldots] \,=\, \int_{F_{\mu\nu}\,=\,0}  [DA_\mu][\ldots]\,.
\ee

Suppose we puncture~$S_1\times S^{\tau}_3$ by excising the~$T_2$ defined by the constraint~$\theta=0$ on~\eqref{background_metric00Intro}. Then, at least formally, one can define the path integral of~$e^{-\mathcal{Q}V}$ with fixed boundary conditions for the tangential components of the gauge field~$A_\mu$ at~$\theta=0$: let us call these boundary conditions~$A_\gamma$. As for every other field at the puncture, including the normal components of the four-dimensional gauge field~$A_\mu$, one integrates over them. The result of such a path integral must be a (two-dimensional) gauge invariant functional of~$A_\gamma$
\be\label{EffectiveAction}
\mathcal{I}^{\times}\,:=\,e^{-S^{\times}_{\text{2d-\,}\text{eff}}[A_\gamma]} \,.
\ee
with a Schwinger-Dyson quantum effective action~$S^{\times}_{\text{2d-\,}\text{eff}}\,$.~\footnote{The contributions coming from the gauge-fixing Fadeev-Popov determinant associated with the four-dimensional gauge degeneracy can be understood to be part of the effective action. } This action needs not to be local, and indeed in the process of integrating out other degrees of freedom, determinant contributions for~$A_\gamma$ can arise, for instance, from integrating out charged matter fields (as it is well known to happen in closely related examples~\cite{Gerasimov:2006zt}). 

Formally
\be\label{RelationFunctional}
\mathcal{I} \=\int \frac{D[A_\gamma]}{\text{vol(orbit $A_\gamma$)}}\,\mathcal{I}^{\times} \,,
\ee
where by vol(orbit $A_\gamma$), we denote the volume of the orbits induced by the action of the four-dimensional gauge transformations upon the $2d$ connection~$A_\gamma$.

The localization formula~\eqref{LocalizationPI} implies that
\be\label{Localization2d}
\mathcal{I}\,=\,\int \frac{D[A_\gamma]}{\text{vol(orbit $A_\gamma$)}}\,\mathcal{I}^{\times} \,=\, \int_{F_{12}[A_\gamma]=0} \frac{D[A_\gamma]}{\text{vol(orbit $A_\gamma$)}}\,\mathcal{I}^{\times} 
\ee
where~$F_{12}$ denote the components of the field strength tensor along the excised~$T_2$.
Thus:
\begin{itemize}
 \item { The localization method can be used to reduce the path integral~\eqref{RelationFunctional} to an integral over the moduli space of two-dimensional flat connections~$A_\gamma$.}
 \end{itemize}
 {\label{FootnoteLocalization} More precisely, the localization method constraints the effective action~$S^{\times}_{\text{2d-\,}\text{eff}}[A_\gamma]$ to be a closed element (not necessarily exact) in the equivariant cohomology of the moduli space of flat connections on~$T_2$. 
 
 Applying non-abelian equivariant localization~\footnote{A review of this method can be found in~\cite{Cordes:1994fc}.} one could expect to further localize~\eqref{Localization2d} to a sum over a disconnected set of families of flat connections~$L$. } Indeed, starting from a matrix integral representation that will be reviewed below, the index can be reduced to a form~\cite{Closset:2017bse}\cite{Benini:2018mlo}\cite{Cabo-Bizet:2020ewf} -- that we schematically represent as --
  \be\label{BetheAnsatzRep}
 \mathcal{I}\= \sum_{L} \frac{\mathcal{I}^{\times}(L)}{\mathcal{H}_{1-loop}(L)}\,.
 \ee
\footnote{Some of these~$L$'s can be continuous families of solutions, and integration over such a subfamily must be considered. To keep our presentation as simple as possible we are ignoring this subtlety. } These~$L$'s are called Bethe roots or fixed points of an equivariant action. The positions of these fixed-points are defined by Bethe ansatz-like equations. The~$\mathcal{H}_{1-loop}$ relates to the determinant of the kinetic operator of a localization-exact deformation of the effective action in~$S^{\times}_{\text{2d-\,}\text{eff}}\,$. 
 
 In conclusion, the path integral~\eqref{RelationFunctional} -- which in principle runs over generic regular field configurations~${A}_\gamma$ -- localizes to a path integral over the moduli space of 2d flat connections~\cite{Cordes:1994fc}~
 \be\label{GaugeFlat}
 {A}_\gamma =  -\text{i} G_{\times}^{-1} d_\gamma G_{\times} \,.
 \ee
 Such an integral can be always recast in the form
 \be\label{Reduction}
 \mathcal{I}\,=\, \int \frac{D[G_{\times}]}{(vol(\text{trivial~$G_{\times}$}))}\,\,  e^{-W_1(G_{\times})}
 \ee
with~$G_{\times}$ being a regular map from the torus~$T_2$ to the group~$SU(N)$. By $vol(\text{trivial~$G_{\times}$})$ we mean the volume of the space of maps with trivial holonomy.

The interesting problem is not to write down the formal expression~\eqref{Reduction}, but to determine the correct effective action~$W_1$ that, again, needs not be local. Of course, further cohomological reductions must be possible on~\eqref{Reduction}, as it is already known to be localizable to the Bethe ansatz form~\eqref{BetheAnsatzRep}.

So far, we have discussed three different \emph{lower-dimensional} representations of the superconformal index,~\eqref{Localization2d},~\eqref{BetheAnsatzRep},~\eqref{Reduction}, the first and the last one, formal, the second one, as we will recall below is entirely explicit. The three of them will be helpful next.

\paragraph{Bootstrapping the two-dimensional effective theory at~$\times$} 
Let us focus on the representation~\eqref{Localization2d}.

The gauge covariant data in a flat connection~$A_\gamma$ is encoded in path ordered holonomies
\be\label{HolonomyField}
M_\ell \,:=\, P e^{ \int_\ell  A_{\gamma=\ell}}\,,
\ee
where~$\ell$ is a generic (space-like) one-cycle in~$T_2\,$ and~$A_{\gamma=\ell}=:A_\ell$ denote the component of~$A_\gamma$ along the cycle~$\ell$.  Formally, the set of OPEs defining the effective theory at~$\times$ is encoded~\footnote{... by using a map that in principle depends on the two spacetime coordinates in~$\times\,$.} in the factorization properties of the covariant correlators
\be\label{CorrelatorsM}
\int \frac{D[A_\gamma]}{\text{vol(orbit $A_\gamma$)}}\,e^{-S^{\times}_{\text{2d-\,}\text{eff}}[A_\gamma]}\,  \prod_{\textbf{r}}\,M_{\ell}[A_\gamma]\,,
\ee
where~$\textbf{r}$ denote irreducible representation of~$SU(N)$.
 
 The gauge invariant data is encoded in traces of the correlators~\eqref{CorrelatorsM}~\footnote{Notice that products involving monodromy operators along disconnected loops ${\ell_{i=1,\ldots p}}$ can be always understood as a monodromy operator along a connected loop with pieces of the contour cancelling pairwise. We are not demanding the effective theory to be topological.}
\be\label{CorrelatorsMTr}
\int \frac{D[A_\gamma]}{\text{vol(orbit $A_\gamma$)}}\,e^{-S^{\times}_{\text{2d-\,}\text{eff}}[A_\gamma]}\,  (\underset{\textbf{r}_1}{\text{Tr}}\,M_{\ell}[A_\gamma])\,\ldots\, (\underset{\textbf{r}_{n}}{\text{Tr}}\,M_{\ell}[A_\gamma])
\ee

If one infers a product algebra (OPE) associated to the factorization of the covariant correlators~\eqref{CorrelatorsM} (assuming such a factorization exists) then one could use that to constrain the possible two-dimensional effective theories at the 2-torus puncture (up to dualities). 

\subsection{A universal quantum group structure from gauge invariance}\label{sec:QuantumGroup}

Classical results~\cite{Alvarez-Gaume:1988bek,Alekseev:1992wn}\cite{Verlinde:1988sn,Moore:1988qv,Moore:1989ni}, which essentially rely only on the algebraic structure~\eqref{KMalgebraText} below~\cite{Alekseev:1992wn}, imply that~\emph{the subset of correlators~\eqref{CorrelatorsM} for a fixed cycle~$\ell$ factorizes as follows from the fusion rules of representations of~$U_q(SU(N))$. }~\footnote{A review of the results of~\cite{Alvarez-Gaume:1988bek,Alekseev:1992wn} is left for future work. In this manuscript we are simply borrowing their results.}  

In Appendix~\ref{SUNKM} we show that $SU(N)_1$ KM algebra 
\be\label{KMalgebraText}
-\i [A^a_{\ell}(\alpha), A^b_{\ell}(\alpha^\prime)]\= -\, \delta^{a b}\,\delta_{per}^{\prime}(\alpha-\alpha^\prime) \,+\,\delta_{per}(\alpha-
\alpha^\prime)\,f^{a b}_{\,\,\,\, c} A^c_\ell(\alpha)\,,
\ee
is the global current algebra induced by the four-dimensional gauge transformations over the phase space of tangential component~$A_{\gamma=\ell}=\sum_{a} A^{a}_{\ell} \,X_a$ of a flat connections along a cycle~$\gamma=\ell$ -- with worldline coordinate~$\alpha\sim \alpha+2\pi$ -- at the punctured boundary~$\times$. The~$X_a$ are a basis of matrices for the adjoint representation of~$SU(N)$. As~$\mathcal{I}$ and~$S^{\times}_{\text{eff}}$ must be invariant under the transformations induced by four-dimensional gauge transformations then the space of modes over which the effective measure~$\int {D[A_\gamma=-\text{i}G_\times^{-1}d_\gamma G_\times]}$ is defined over --must split in representations of such~$SU(N)_1$ KM algebra.

The~$SU(N)_{1}$ KM algebra carried by the~$A_\ell$'s is known to induce a~$U_q(SU(N))$ quantum group structure in the OPE among holonomy variables~$M_\ell$~\cite{Alvarez-Gaume:1988bek,Alekseev:1992wn}\cite{Verlinde:1988sn,Moore:1988qv,Moore:1989ni} with $q=\exp{\frac{\pi\text{i}}{1+N}}$. Relying on the results of~\cite{Alekseev:1992wn}, the monodromy matrices~$M_\ell$ are bound to behave as~$T$-matrices of an underlying integrable system, in the sense that they satisfy the so-called~$RTT$ relations with respect to the~$R$-matrix of~$U_q(SU(N))$, more precisely the~$RTT$ relations are known to be equivalent to the defining commutation relations of~$U_q(SU(N))$~\cite{Alekseev:1992wn}.

The~$A_\ell$'s that are related by trivial four-dimensional gauge transformations,~\footnote{Represented as vol(\text{Trivial~$G_{\times}$}) in \eqref{Reduction}, where by trivial we mean that the corresponding map generates a connection with trivial monodromy along every one-cycle~$\ell$. } form irreducible representations of the affine Kac-Moody algebra~\eqref{KMalgebraText}. Single gauge orbits are classified by a gauge-invariant observable, which should depend on the irreducible representation. Irreducible representations of both the affine Kac-Moody algebra and~$U_q(SU(N))$ can be labelled by the labels of their highest weight state, which we schematically represent with the letter~$\textbf{r}\,$. The monodromy operators give a natural gauge-invariant observable
\be
\mathcal{O}_{\textbf{r}}[A_\ell]:=\underset{\textbf{r}}{\text{Tr}} M_{\ell}[A_\gamma]\,.
\ee
\newcommand{\Op}{$\mathcal{O}_{\textbf{r}}[A_\gamma]$}
 Moreover, as only states that belong to the same irreducible representation~$\textbf{r}$ are related by trivial gauge transformations, we conclude that:
 \begin{itemize}
 \item  The localized gauge orbits~$L$ (Bethe roots) are in one-to-one relation with irreps~$\textbf{r}$ of the quantum group~$U_q(SU(N))$.
 \end{itemize}

This implies that in a limit of~$\mathcal{I}$ where only~$N_0$ localized gauge orbits remain un-suppressed
\be\label{ConstraintRep}
N_0 \=\text{a~\# of irreps \textbf{r} of } U_q(SU(N))\,.
\ee

Now, what is equation~\eqref{ConstraintRep} useful for?

The OPE among ``gauge orbits"~$L$ -- represented by the operators~\Op -- is fixed by the $U_{q}(SU(N))$ structure (more precisely by the so-called co-product structure, which will be not reviewed in this manuscript). The finite set of~$L$'s that dominate the corresponding limit of~$\mathcal{I}\,$, must be closed under the previously mentioned OPE structure, otherwise the quantum group symmetry would be broken and consequently the four-dimensional gauge theory would be anomalous, which we know it is not the case.~\footnote{This OPE's must match the answers obtained from~\eqref{CorrelatorsMTr} in the corresponding limit. As far as we understand, there are not many of such possible closed~$OPE$-structures.} 

There exists a unique set of irreps of~$U_{q}(SU(N))$ which is closed under the OPE structure above mentioned, it is the so-called set of integrable representations. For~$q=e^{\frac{\pi\text{i}}{1+N}}$ such a set is composed of~$N$ elements -- the corresponding integrable representations -- which are known to be counted by the partition function of~$SU(N)_1/SU(N)_1$ WZNW model on~$T_2\,$, a~TCFT$_2$.

 Of course, one can always consider direct sums of such OPE-closed set and that implies that the number of gauge orbits~$N_0$ must be a multiple of~$N$, i.e., that
\be\label{DivisorConstraint}
N_0| N\,,
\ee 
which is in a sense a sharper statement than~\eqref{DivisorConstraint}.

For example, in the limit $\tau \to 0$ where~$N_0=N\,$, the constraint~\eqref{DivisorConstraint} implies that the two-dimensional effective theory at the puncture must be equivalent (or dual) to the gauged~$SU(N)_1/SU(N)_1$ WZNW model on~$T_2$.
 
The previous analysis implies also other partial conclusions: 
\begin{itemize}

\item To each gauge orbit~$L$ one can associate one of the~$N$ fundamental representations of~$SU(N)$ (the antisymmetric tensor products of the defining~$N$-dimensional representation) which we denote with the letter~$\textbf{r}\,$, and a gauge invariant non-local (line) excitation/operator~$\mathcal{O}_{\textbf{r}}[A_\mu]$ at a fixed Cauchy surface (line)~$\ell$. 

\item The latter~$N$ gauge orbits~$L$ are in one-to-one relation with the~$N$ (chiral) conformal blocks of the parent (ungauged) WZNW model: which are the degrees freedom counted by the partition function of the gauged WZNW model~\cite{Karabali:1989dk,Spiegelglas:1992jg}~\cite{Gawedzki:1988nj,Witten:1991mm}. {~\footnote{In particular, the conformal primaries of the parent(ungauged) WZNW are also in one-to-one relation with the latter gauge orbits~$L$. The conformal spectrum of this CFT$_2$~\cite{Witten:1983ar,Polyakov:1983tt} is defined by the one of the integrable or highest weight representations of its chiral affine algebras~\eqref{KMalgebraText}~\cite{Bernard:1987df} , -- that is because the corresponding mass matrix is diagonal--. Thus, the conformal primaries of this theory are in one-to-one relation with the integrable representations of the chiral affine algebras~\eqref{KMalgebraText}~\cite{Bernard:1987df}, and of those of the underlying~$U_{q}(SU(N))\,$}}

\item {The non-local operators~$\mathcal{O}_{\textbf{r}}$ inherit an OPE structure (at fixed time) which is equivalent to the Verlinde fusion algebra~\cite{Verlinde:1988sn} of the underlying~$U_q(SU(N))$. Thus, in physical terms, the corresponding excitations are a one-dimensional higher generalization of three-dimensional Anyons~\cite{Moore:1991ks,Read:1991bt}~\cite{2008RvMP...80.1083N}.  The worldvolume spanned by the propagation in time of the excitations created by the operator~$\mathcal{O}_{\textbf{r}}$ in the Hilbert space of the four-dimensional gauge theory at a fixed Cauchy surface,  is not a line --  as it is the case for Anyons propagating in three-dimensions and more generally for point-like particles propagating in any dimension -- but a surface. Thus, the corresponding excitation is non-local, more precisely one-dimensional at a fixed-Cauchy surface.}

\end{itemize}

 \subsection{Another perspective on the gauge orbits~$L$: The fibering operator and condensates }\label{sec:FiberingOperator}

This subsection presents a different approach. The superconformal index~$\mathcal{I}$ can be written as an expectation value of a surface-operator~$\mathcal{F}$ in a different supersymmetric twist of~$\mathcal{N}=4$ SYM when the theory is placed on~$T_2\times S_2$~\cite{Closset:2017bse}
\be\label{ATwistExp}
\mathcal{I}\= \left<\,\mathcal{F}\,\right>_{T_2\times S_2}\,.
\ee
The~$\mathcal{F}$ is the so-called~\emph{fibering operator}~\cite{Closset:2017bse}\cite{Closset:2017zgf}. This is a surface operator that wraps the~$T_2\,$. The~$\left<\ldots\right>_{\ldots}$ means expectation value in the four-dimensional~$A$-twist of~$\mathcal{N}=4$ SYM on~$T_2\times S_2$
\be\label{MetricT2S2}
ds^2 \,=\, d{t}_A^2 \,+\, (\text{d}{\phi}_A\, +\, \tau \text{d}t_A)^2\,+\,(\text{d}{\theta_A})^2\,+\,\sin^2{\theta_A} (\text{d}\psi_A)^2
\ee
with~$\theta_A\in [0,\pi]$ and~$\phi_A,\chi_A\in [0,2\pi]$ and~$t_A\sim t_A+\beta\,$.

Let us briefly review the approach of~\cite{Closset:2017bse}\cite{Closset:2017zgf}. After using supersymmetric localization to compute the expectation value in the right-hand side of~\eqref{ATwistExp}~\footnote{... not to be confused with the twisted~$S_1\times S_3$ partition function in the right-hand side of~\eqref{PathIntegral} that instead localizes to~\eqref{TheIndex}~\cite{Nawata:2011un}\cite{Cabo-Bizet:2018ehj}.} the authors found that
\be\label{VEVFT2S2}
\left<\,\mathcal{F}\,\right>_{T_2\times S_2}\= \sum_L\frac{\mathcal{F}(L)}{\mathcal{H}(L)}\,,
\ee
where~$\mathcal{F}(L)$ equals the integrand of~\eqref{TheIndex} evaluated at a specific set of configurations~$L$. These configurations correspond to vacua of  the dimensionally reduced A-twist (on~$T_2$)~\cite{Closset:2017bse}\cite{Closset:2017zgf}, namely they correspond to~$SU(N)$ flat connections on~$T_2$. These connections are parameterized by~$N-1$ complex variables~$v_a\,$, and they are subject to the identifications 
 \be
 v_a\, \sim\, v_a \,+\,1\,, \qquad  v_a\,\sim\, v_a \,+\, \tau\,.
 \ee
 The positions~$v_a$ of the vacua~$\{L\}$ are determined by a Bethe ansatz like equation
\be
\frac{Q_a}{Q_N}\,:=\,e^{-\frac{\partial \mathcal{W}}{\partial v_a}}=1\,, \qquad a\,=\,1\,,\ldots\,, N\,-\,1\,.
\ee
which we write/solve (in Cardy-like limit) below in equation~\eqref{BetheAsymptotics}. The~$Q_a\,$, $a=1,\ldots,N-1$ are called Bethe ansatz operators (the definition of~$Q_a$ for~$SU(N)$ $\mathcal{N}=4$ SYM is recalled in~\eqref{BetheOperator}). The fixed-points~\footnote{ These are saddles of the twisted superpotential~$\mathcal{W}$. The fibering operator $\mathcal{F}$ and~$\mathcal{H}$ are defined in terms of~$\mathcal{W}$. {These definitions, which are not relevant to our discussion below, can be found in the original reference~\cite{Closset:2017bse}.}} relevant to the present discussion are
\be\label{SolsCardyIntro}
v_a\= \frac{a\,-\,\widehat{a}}{N}\, (m \tau\,+\,n)\,,
\ee
with the label~$\widehat{a}\,=\,0,\ldots, N-1$ denoting different solutions~$L$.~\eqref{SolsCardyIntro} is not a gauge invariant characterization of the orbits~$L$. In terms of the~$v_a$'s the gauge-invariant representative~\Op~localizes into character~$\chi_{\textbf{r}}(v)$ of the Lie group~$SU(N)$ for some representation~$\textbf{r}$.  These representations~$\textbf{r}$ are emergent in the sense they need not be just the adjoint, which is the unique representation carried by the fields in the perturbative Lagrangian formulation of the theory.~In this case the irreps~$\{\textbf{r}\}$ denote the finite-dimensional highest weight representations of the Lie algebra~$SU(N)$ that descend from integrable representations of the affine Kac-Moody algebra~\eqref{KMalgebraText} at level~$k=1$, the ones labelled by the~$N$ fundamental weights of~$SU(N)$.~\footnote{These are the finite-dimensional highest weight representations of the simple Lie algebra~$SU(N)$ that comes from the projection to grade~$0$ generators of a given integrable representation of the affine algebra~$SU(N)_{k=1}$.} They are also related to the labels~$\widehat{a}\,$, i.e., different values of~$\widehat{a}$ are in one-to-one relation with the~$N$ fundamental irreps~$\textbf{r}$ of~$SU(N)\,$.

\cite{Closset:2017bse} noticed that the one-loop exact computation of the four-dimensional A-twist observable~\eqref{VEVFT2S2} on~$T_2\times S_2$, can be identified with the one of the superconformal index on~$S_1\times S_3$ after identification of KK modes in the two different manifolds. We will comeback to this identification below. Using this observation the authors conjectured and perturbatively tested~\eqref{VEVFT2S2} in some examples. Later on reference~\cite{Benini:2018mlo} derived the same formula
\be\label{BARepIndex}
\mathcal{I}\= \sum_L\frac{\mathcal{F}(L)}{\mathcal{H}(L)}\,,
\ee
following an entirely independent approach. This approach does not use the relation with the A-twist on~$T_2\times S_2\,$, as it relies only on the quasiperiodicity properties of the analytic extension of the integrand of~\eqref{TheIndex}. More recently, this formula has been understood to be a realization of the Atiyah-Bott-Berline-Vergne equivariant integration formula over a double dimensional space of complexified holonomies.~{This is shown in version 2 of~\cite{Cabo-Bizet:2020ewf} (to appear). In section~\ref{DoubleLimit} we will use this last perspective to revisit the computation of the asymptotic exponential growth of the superconformal index~$\mathcal{I}\,$ in the limits~$\tau\to -\frac{n}{m}$.~\footnote{We note that this approach to the Cardy-like expansion is independent, and thus complementary to, of the one studied in the companion paper~\cite{PaperBulk}. }

Note that~\eqref{BARepIndex} is a concrete realization of the general discussion above given, with the natural identification~$\frac{\mathcal{I}^{\times}(L)}{\mathcal{H}_{\text{1-loop}(L)}}\,\sim\, \frac{\mathcal{F}(L)}{\mathcal{H}(L)}$\,.~\footnote{Note that we do not identify~$\mathcal{H}_{\text{1-loop}}(L) \,=\, \mathcal{H}(L)$, that is because there could be determinants coming from integrating out matter or ghost fields included in the definition of~$\mathcal{I}^{\times}\,$.}
~The formula~\eqref{BARepIndex} is called the Bethe ansatz representation of the superconformal index. The~$L's$ correspond to condensates of the surface fibering operator~$\mathcal{F}$ on the trivially fibered~$T_2\,\subset\,T_2\times S_2$.~\footnote{In a sense the expectation value of the fibering operator at the state~$L$,~$\mathcal{F}(L)$, can be interpreted to be the expectation value of the operator~$\mathcal{O}_{\textbf{r}}$} On~$S_1\times S_3\,$ these condensates translate into topological fluxes localized at two two-torus punctures~$\times_1$ and~$\times_2\,$, i.e., at the worldvolume of~$\mathcal{B}\,$. Let us explain how this happens.

\paragraph{Surface operators and condensates}

The natural identification between the coordinates in~\eqref{MetricT2S2} and the coordinates of a rotating spacetime~\eqref{background_metric0} is
\be
{\theta}_A \,\sim\, 2\theta\,,\,\qquad {\chi}_A \,\sim\,\frac{\phi_1\,-\,\phi_2}{2}\,, \qquad {\phi}_A \,\sim\,   \frac{\phi_1\,+\,\phi_2}{2}\,,\qquad {t}_A \,\sim\, t_E\,.
\ee
Then the non-trivial fibration transforming~\eqref{MetricT2S2} into~\eqref{background_metric00Intro} is implemented by the shift
\be\label{changeHopf}
\text{d}{\phi_A}\,+\,\tau dt_A \,\longrightarrow\, (\text{d}{\phi}_A\,+\,\tau dt_A\,+\,\cos \theta_A  \text{d}\psi_A).
\ee
As mentioned before, with these identifications and transformation~\eqref{changeHopf} the spectrum of eigen-modes contributing to the computation of the A-twisted path integral in the geometry~\eqref{MetricT2S2} is equivalent to the one relevant for the computation of the superconformal twist in the geometry~\eqref{background_metric00Intro}~\cite{Cabo-Bizet:2018ehj}, as observed in~\cite{Closset:2017bse}.

The main partial conclusion we will draw next is that:
\begin{itemize}
\item {A generic smooth gauge connection along the $T^\tau_2 \,\subset\, T^\tau_{2}\times S_2$ corresponds to a singular gauge connection on the four-dimensional space~$S_1\times S^\tau_{3}$. These singularities can be understood as a surface operator~$\mathcal{B}$ that wraps two disconnected~$T_2\text{'s}\,\subset\,S_1\times S^\tau_{3}\,$~\eqref{background_metric00}: }
\end{itemize}
This is because the cycles~$\phi_1$ and~$\phi_2$ that add up to form the cycle~$\phi$, the one that the operator~$\mathcal{F}$ fibers over~$S_2$ to recover~$S_1\times S^{\tau}_3$, are contractible at~$\theta=0$ and $\theta=\frac{\pi}{2}$, respectively. For instance, a gauge connection of the form~$A_{{\phi}_A}\,\neq\,0$ and~$A_{{\chi}_A}\,=\,0$ would translate into a singular gauge connection~$A_{\phi_1}=A_{\phi_2}\,\neq\,0$ on~$S_{1}\times S^{\tau}_3$. In other words, the latter gauge connections define an infinite density of flux localized at the 2-tori~$\theta=0$ and~$\theta=\frac{\pi}{2}$. Section~\ref{subsec:MagElect} shows how this concentration of flux happens for the localized gauge orbits~$L$ that dominate the Cardy-like limit~\eqref{SolsCardyIntro}.
\begin{itemize}
 \item {In this sense, we conclude that inserting the fibering operator~$\mathcal{F}$ in~$T^\tau_2 \,\subset\, T^\tau_{2}\times S_2$ \emph{defines} an operator~$\mathcal{B}$ on~$S_1\times S^\tau_3$. This operator wraps two-disconnected~$T_2$ punctures in~$S_1\times S^\tau_{3}\,$. }
\end{itemize}
The properties of the states on which~$\mathcal{B}$ can condensate are defined by the~$L$'s. In Cardy-like limits we will show that these condensates carry fractional electromagnetic flux and~$\mathbb{Z}^{(1)}_N$ one-form charge (See section~\ref{CondensatesCharges}). 

~The operator~$\mathcal{B}$ represents the deformation (of the Hilbert space of states) implemented by fixing non-trivial boundary conditions~$A^{\times_{1,2}}_\gamma$ at two-punctures on~$S_1\times S^\tau_{3}\,$. The vacuum expectation value~$\bigl<\mathcal{B}\bigr>$ is formally defined as a path integral~$\mathcal{I}^{\times\times}$ in the presence of two insertions,
the obvious generalization of~$\mathcal{I}^{\times}$, but this time with two punctures, one at~$\theta=0\,$, which we denote as~$\times_1\,$, and the other at~$\theta=\frac{\pi}{2}$, $\times_2\,$. By definition
\be\label{TwoPuncturesAction}
\int \frac{D[A^{\times_1}_\gamma]}{\text{vol(orbit $A^{\times_1}_\gamma$)}} \int\frac{D[A^{\times_2}_\gamma]}{\text{vol(orbit $A^{\times_2}_\gamma$)}}\, \,\mathcal{I}^{\times\times} \= \mathcal{I} \,.
\ee
The defining relation of~$\mathcal{B}$ in terms of the fibering operator~$\mathcal{F}$ and the localization method imply
\be
\sum_{L} \frac{\mathcal{I}^{\times\times}(L)}{\mathcal{H}_{one-loop} (L)}\=\sum_L \frac{\mathcal{F}(L)}{\mathcal{H}(L)}\,.
\ee
Repeating the same reasoning as before but starting from the left-hand side of~\eqref{TwoPuncturesAction} one reaches a representation of the form
\be\label{RelationFunctionalTwoP}
\begin{split}
\mathcal{I}& \,=\,
\int\,\frac{{D}G_{\times_1}}{\text{vol}(\text{trivial } G_{\times_1})}\int \frac{{D}G_{\times_2}}{\text{vol}(\text{trivial } G_{\times_2})}\,  e^{-W_2(G_{\times_1},G_{\times_2})}\,.
\end{split}
\ee
Again, in virtue of the localization method, the integral in the left-hand side of~\eqref{TwoPuncturesAction} localizes to an integral over two-dimensional flat connections~$A^{\times_{1,2}}_\gamma$ on the tori at~$\times_{1,2}$, which can be written as an integral over regular maps~$G_{\times_1\times_2}$ (with non trivial holonomy) from~$T_2$ to~$SU(N)$.

In Cardy-like limit the effective theory defined by~$W_2$, must reduce to G/G WZNW model with~$G=SU(N)_1$ on~$T_2$. This will then constrain the form of~$W_2$ drastically as it is explained in appendix~\ref{sec:WZNW}.

\paragraph{Relation to the remaining part of the paper}

We should briefly comment about how the analysis presented in this section relates to the the scope of the remaining part of this paper.

Section~\ref{sec2} introduces background material that has been already reported in the literature. The scope of that section is to review more details of the localization computations that were summarized before in this section. Sections~\ref{CondensatesCharges} and~\ref{subsec:MagElect} show that the gauge orbits~$L$ correspond to localized Dyonic flux at the~$T_2$ punctures $\theta=0$ and~$\theta=\frac{\pi}{2}$, namely it studies the condensates of the surface operator~$\mathcal{B}$ at the relevant Bethe vacua~$L$'s.  Finally, subsection~\ref{OneFormSymmetry} shows how  -- as for their analogous system of Abelian Anyons in three-dimensions -- the condensates of the gauge orbits~$L$ that dominate in Cardy-like limit form a reprentation of a $\mathbb{Z}^{(1)}_N$ one-form symmetry. 

\section{Quantum phases of~4d~$SU(N)$~$\mathcal{N}=4$ SYM}
\label{sec2}
This section gives a pedagogical introduction to necessary and known background material. 

\subsection{Zero-temperature BPS limit}
\label{DoubleLimit}

Four dimensional~$SU(N)$~$\mathcal{N}=4$ SYM has a~$PSU(2,2|4)\,$ global symmetry algebra. Among the even Cartan generators one finds energy~$E$, the two spin-1/2 Cartan generators~$J_1=\frac{1}{2}(J_{3L}+J_{3R})$ and~$J_2=\frac{1}{2}(J_{3L}-J_{3R})$ of~$so(4)\,=\,su(2)_{J_{3L}}\,\oplus\, su(2)_{J_{2R}}$, and the Cartan generators~$R_1$,~$R_2$,~$R_3$ of the R-symmetry algebra~$su(4)$, in some conventions. It is always possible to find two complex conjugated supercharges~$Q$ and~$Q^\dagger\=S$ in~$PSU(2,2|4)$ such that
\be\label{BPSConditionN4SYM}
2\{Q,S\}\= \widetilde{H}\= E-J_1-J_2-R_1-R_2-R_3\,\geq\,0\,.
\ee
Let~$\X$ be the set of~$\frac{1}{16}$ BPS states in the Hilbert space~$\mathcal{H}$ that are anihilated by both~$Q$ and~$S$. The charges of states in~$\X$ saturate the BPS condition~\eqref{BPSConditionN4SYM}. Let~$F$ be the fermionic number associated to~$Q$ and~$S\,$. Assume that for any state not belonging to~$\X$,~
\be\label{Gap}
\widetilde{H}\,\geq \Delta\,>\,0\,.
\ee
$\Delta$ will be called gap.

Define the chemical potentials dual to~$J_1\,$,~$J_2\,$, and~$R\equiv R_1 +R_2+R_3$ as
\be
\Omega_{1}\,\equiv\, \frac{\omega_1+\beta}{\beta}\,,\,\quad
\Omega_{2}\,\equiv\, \frac{\omega_2+\beta}{\beta}\,,\,\quad \Phi\,\equiv\, \frac{\varphi+\beta}{\beta}\,.
\ee
Given~\eqref{BPSConditionN4SYM} and~\eqref{Gap}, the zero-temperature limit~\cite{Cabo-Bizet:2018ehj}
\be\label{BPSLimit1}
\begin{split}
\beta &\,\to\, \infty\,, \\\omega_{1,2}&\,\to\,\text{finite}\,,\\ \omega_1\,+\,\omega_2\,-\,2\varphi&\,\equiv\, 2\mu\=\pm\,2\pi\i\,+\, \mu^{(1)}(\beta)\,\to\, 2\pi\i\,,
\end{split}
\ee
of the partition function~
\be\label{PFunction}
{Z}(\beta,\,\mu\,,\omega_{1,2})\,\equiv\, \underset{\mathcal{H}}{\text{Tr}}\,e^{-\beta H}\,e^{\beta\Omega_1 J_1\,+\,\beta\Omega_2 J_2 \,+\,\beta\Phi R} \= \underset{\mathcal{H}}{\text{Tr}}\,e^{-\beta \widetilde{H}}\,e^{\omega_1 J_1\,+\,\omega_2 J_2 \,+\,\varphi R} \,,
\ee
reduces to a~$(-1)^F$ graded trace~\footnote{Other gradings, not as protected as the superconformal index, have been recently studied in~\cite{Cherman:2020zea}.}, as follows from the algebraic manipulations
\be\label{Steps}
\begin{split}
\underset{\X}{\text{Tr}}\,e^{\omega_1 J_1\,+\,\omega_2 J_2 \,+\,\varphi R} \= \underset{\X}{\text{Tr}}\,e^{\mu F}e^{\,(\omega_1\,-\,2\mu) \widetilde{J}_1\,+\,\omega_2 \widetilde{J}_2}&\=\underset{\mathcal{H}}{\text{Tr}}\,(-1)^F \,e^{\,(\omega_1\,-\,2\mu) \widetilde{J}_1\,+\,\omega_2 \widetilde{J}_2}\\
&\,\equiv\,\mathcal{I}(\omega_1-2\mu, \omega_2)\,, \qquad \mu^{(1)}(\beta)\,=\,0\,,
\end{split}
\ee
where~$\widetilde{J}_1=J_1+\frac{R}{2}\,$,~$\widetilde{J}_2=J_2+\frac{R}{2}\,$. In~\eqref{Steps} we have assumed the spin-statistics relation~$F= 2 J_1 \text{mod}\mathbb{Z}\,$.

$\mathcal{I}$ is the superconformal index of~$SU(N)$ $\mathcal{N}=4$ SYM, and it can be recast as a~$(N-1)$-dimensional integral~\cite{Romelsberger:2005eg,Kinney:2005ej,Dolan:2008qi}
\be\label{N4SYM0}
\mathcal{I}\=\int^1_0 \prod^{N\,-\,1}_{i\,=\,1} dv_a \, e^{\mathcal{V}}\,,
\ee
with a periodic potential~$\mathcal{V}\,$. The~$v_a\,\sim\,v_a+1$ will be called~\emph{eigenvalues}.~\footnote{They can be thought of as eigenvalues of a traceless~$N\times N$ Hermitian matrix.}  The physical meaning of the~$v_a$'s, which will be introduced in the next section, will be essential in our discussion.

At any temperature, choosing~$\mu^{(1)}=0$ reduces~$Z$ to~$\mathcal{I}\,$, and  the temperature dependence disappears. For~$\mu^{(1)}\,\neq\, 0$, but still~$\mu^{(1)}\,\underset{\beta\to \infty}{\rightarrow}\, 0$, there are no cancellations at finite temperatures, and only at~$\beta\=\infty$~$Z$ reduces to~$\mathcal{I}$. The first small-temperature correction (not coming from the implicit dependence of~$\mu^{(1)}$ on~$\beta$) to the BPS limit~\eqref{BPSLimit1} of~$Z\,$, is suppressed with the gap~$\Delta$~\footnote{Note that this correction is independent of the specific value of~$\mu^{(1)}$, as long as the latter is different from zero. } i.e it is of order~$\sim e^{-\beta\Delta}$. Assuming~$\mu^{(1)}$ tends to zero fast enough as~$\beta\to \infty$ implies that for small enough temperatures, the temperature-dependent corrections to the index coming from the implicit dependence in~$\mu^{(1)}$ can be neglected with respect to the contributions coming from the gap at order~$e^{-\beta \Delta}\,$.

In the gravitational BPS limits of~\cite{Cabo-Bizet:2018ehj,Cassani:2019mms} the chemical potentials depend implicitly on the Bekenstein-Hawking temperature. The small-temperature corrections to the BPS onshell action come from the implicit dependence of~$\omega_{1,2}$ and~$\mu\,$ on the Bekenstein-Hawking temperature.  These classical corrections seem to be conceptually unrelated to the order~$e^{-\beta\Delta}$ corrections mentioned in the previous paragraph. By ignoring the dependence of~$\omega_{1,2}$ and~$\mu$ on temperature, we have chosen not to pay attention to the former kind of corrections and to their corresponding zero-temperature limits~\footnote{These are the semiclassical limits studied, for instance, in~\cite{Larsen:2019oll}. Interestingly, reference~\cite{Heydeman:2020hhw} has argued that perturbative quantum corrections in the bulk can produce exponentially suppressed contributions, that could be related to the gap~$\Delta\,$.}, which will be analyzed elsewhere.

\subsection{The index as a path integral: The twisted time cycle}\label{SUSYLocalization}

The superconformal index~$\mathcal{I}$ can also be understood as a supersymmetric path integral of~$4d$ $\mathcal{N}=4$ SYM on a non trivial fibration of~$S_3$ over~$S_1\,$ with appropriate periodicity conditions for bosons and fermions along a twisted time-cycle of a given backrgound geometry~\cite{Cabo-Bizet:2020ewf}.
In such a geometry, as one moves along the~$S_1$, the~$S_3$ rotates around two independent directions. The corresponding angular velocities can be related to the parameters~$\tau$ and~$\sigma$ introduced before. The metric can be taken to be~\cite{Cabo-Bizet:2018ehj} 
\be\label{background_metric00}
\diff s^2 \=  \diff t_E^2 +  \diff\theta^2 + \sin^2 \theta\, \bigl(\diff \phi_1-\i\Omega_1 \diff t_E\bigr)^2 + \cos^2 \theta \,\bigl(\diff \phi_2 -\i\Omega_2 \diff  t_E\bigr)^2 \,,
\ee
where~$\theta\in [0,\frac{\pi}{2}]$ and
\be\label{RelationChemicalPotentials}
\begin{split}
\beta\Omega_{1}&\,=\,\omega_1 \,+\,\beta\,=\,2\pi\i \tau\,+\,\beta\,(1\,+\,\eta_1)\,,\\ \beta\Omega_{2}&\,=\,\omega_2\,+\,\beta\,=\,2\pi\i \sigma\,+\,\beta\,(1\,+\,\eta_2)\,.
\end{split}
\ee
The localization method of~\cite{Pestun:2007rz} can be used to compute the path integral~\eqref{PathIntegral}~\cite{Festuccia:2011ws,Closset:2013sxa,Assel:2014paa}.  See for instance Section 4 of~\cite{Cabo-Bizet:2018ehj} where the method was used with the explicit form of the metric~\eqref{background_metric0} and a background~$U(1)$ graviphoton potential determined by a single parameter~$\Phi$ (as reported around equation 4.1 in that reference). The Killing spinor solution found in~\cite{Cabo-Bizet:2018ehj} depends solely on the combinations of the parameters
\be
\Omega_1\,+\,\Omega_2\,-\,2\Phi
\ee
which means that two backgrounds~$(\Omega_1,\Omega_2,\Phi)$ and~$(\Omega^\prime_1,\Omega^\prime_2,\Phi^\prime)$ such that
\be
\Omega_1\=\Omega_1^\prime\,+\,\beta\, \eta_1\,,\,\Omega_2\=\Omega_2^\prime\,+\,\beta \,\eta_2\,,\,\Phi\=\Phi^{\prime}\,-\,\frac{\beta\,(\eta_1+\eta_2)}{2}\,,
\ee
preserve the same supersymmetry. 

This implies that for any~$\eta_1$ and~$\eta_2$ in~\eqref{RelationChemicalPotentials}, the result of the localization computation can be recovered from the one reported in Section 4 of~\cite{Cabo-Bizet:2018ehj} by substituting
\be\label{substitution}
\omega_1\to 2\pi\i \tau\,,\,\,\omega_2\to 2\pi\i \sigma\,,\,\, \varphi\,\to \,\varphi\,
\ee
in the equations there given.

Focus on the choice~$\eta_1\=\eta_2=-1\,$, then
\be\label{background_metric0}
\diff s^2 \=  \diff t_E^2 +  \diff\theta^2 + \sin^2 \theta\, \bigl(\diff \phi_1+\frac{2\pi\tau}{\beta} \diff t_E\bigr)^2 + \cos^2 \theta \,\bigl(\diff \phi_2 +\frac{2\pi\sigma}{\beta} \diff  t_E\bigr)^2 \,.
\ee
The global structure of the manifold associated to~\eqref{background_metric0}\,, is defined by the periodic identifications
\be\label{GlobalProps}
\phi_1 \sim \phi_1 + 2 \pi\,,\,\phi_2 \sim \phi_2 + 2 \pi\,,\,t_E\,\sim \, t_E\,+\,\beta\,.
\ee
Note that the cycle~$\phi_1\sim\phi_1+2\pi$ contracts at the north~$\theta=0$, and the cycle~$\phi_2\sim \phi_2+2\pi$ contracts at the south ($\theta\=\frac{\pi}{2}$). At the level of the metric, the angular velocities will be assumed to be real as we do not want to sacrifice the semi-positive definiteness of~\eqref{background_metric0} at early stages.~\footnote{A similar argument has bee recently used in the gravitational context~\cite{Iliesiu:2021are}.} Only at advance stages, for instance, in partition functions, or in the superconformal index, analytic extension is useful as a regularization mean.~\footnote{ From the Euclidean path integral perspective, all indications are that placing the $\mathcal{N}=4$ SYM on the analytic extension of the Euclidean metric~\eqref{background_metric0} is fine e.g. the supersymmetric partition function of the complex background geometry is simply an analytic continuation of the answer obtained from the supersymmetric trace formula~\eqref{Steps}, which is obtained by counting supersymmetric operators in the Hilbert space of the Lorentzian theory. As recently done for various examples~\cite{Witten:2021nzp}, an interesting question to ask is if these geometries are allowable accordingly to the axioms of~\cite{Kontsevich:2021dmb}.}

Apart from the obvious time-like and space-like cycles, there is a family of one-cycles that will be relevant in our discussion. From the semi-positive definiteness of~\eqref{background_metric0} and the identification~$t_E\sim t_E+{\beta}$, it follows that~$x^\mu_{\Gamma}(0)\sim x^\mu_{\Gamma}(\beta)\,$ for
\begin{equation}\label{Polyakov0}
x^\mu_{\Gamma}(\a)\,\equiv\,( t_E,\ \theta, \,\phi_1, \phi_2)\=\bigl(\a,\,\,\theta_0\,,\,-\frac{2 \pi \tau}{\beta}\, \a,\,\,-\,\frac{2 \pi \sigma}{\beta}\, \a \bigr)\,, \qquad 0\leq \a <\beta\,.
\end{equation}
We will call these cycles the \emph{twisted time cycles}.
The~$\Gamma$'s can be located at any~$\theta=\theta_0$ in the fiber~$S_3\,$. Note that~$\Gamma$ is, roughly speaking, the~$S_1$ base of the fibration. 

The~$\tau$ (resp.~$\sigma$)-rotation has a fixed locus, which is a rotating two-torus~$(t_{E0},\phi_{20})$~$\Bigl($resp. $(t_{E0},\phi_{10})\Bigr)$ fixed by the condition
\be\label{DiffEqsFirst}
\theta\=0\,\,\,\text{(resp.~$\frac{\pi}{2}$)}\,,\, \quad \frac{\diff \phi_{2}}{\diff \alpha}\=-\,\frac{2\pi \sigma}{\beta} \,\,\,\text{$\Bigl($resp.~$\frac{\diff \phi_{1}}{\diff \alpha}\=-\,\frac{2\pi \tau}{\beta}\Bigr)$}\,,\,\frac{\diff t_E}{\diff \alpha}\=1\,,
\ee
with~$\alpha$ running from~$0$ to~$\beta$. The coordinates along the torus~$t_{E0}$ and~$\phi_{20}$ (resp.~$\phi_{10}$) range over the domain of initial conditions in the first order differential equations~\eqref{DiffEqsFirst}. 

\subsection{Decomposition of~$\mathcal{I}$ in sum over fixed-points: Cardy-like limits }
\label{sec:CardyLimit}
This subsection reviews what is known about the eigenvalue configurations that define the BPS phases~\cite{Cabo-Bizet:2018ehj} following the perspective of~\cite{Cabo-Bizet:2020ewf}.

The integral~\eqref{N4SYM0} can be written as
\be\label{N4SYM}
\mathcal{I}(q)\=\int^1_0 \prod^{N\,-\,1}_{i\,=\,1} dv_a \, e^{\mathcal{V}}\=\int^1_0 \prod^{N\,-\,1}_{i\,=\,1} dv_{1a} \, e^{\mathcal{V}(v_1,v_2=0)}\,,
\ee
with
\be
v_a\= v_{1a}\,+\, v_{2a}\, \tau\,, \qquad v_{1a}\,,\, v_{2a}\,\in\,\mathbb{R}\,.
\ee
Without loss of generality, the potential~$\mathcal{V}$ can be assumed to be doubly periodic~\cite{Cabo-Bizet:2019eaf,Cabo-Bizet:2020ewf}
\be
\mathcal{V}(v_1,v_2)\=\mathcal{V}(v_1+1,v_2)\=\mathcal{V}(v_1,v_2+1)\,.
\ee
More details about~$\mathcal{V}$ can be found in~\cite{Cabo-Bizet:2020ewf}.~\footnote{ We only note that it is related to a smoothening of a periodic extension of the twisted superpotential~$\mathcal{W}$. More details on this can be found in~\cite{Cabo-Bizet:2020ewf}.}
~Exact equivariant integration formulas~\cite{Atiyah:1984px}~\footnote{... which can be proven to be equivalent to~\eqref{BARepIndex}. This will be shown in version 2. of~\cite{Cabo-Bizet:2020ewf}(To appear).
} can be used to determine the asymptotic form of the integral representation~\eqref{TheIndex} at leading order in the Cardy-like expansion~($\tau\to -\frac{n}{m}$ with~gcd$(m,n)=1$) (and finite~$N$). In this way one obtains~\cite{Cabo-Bizet:2020ewf}
\be\label{EquivariantFormula}
\begin{split}
\mathcal{I}&\,\underset{\tau\,\to\,-\frac{n}{m}}{\simeq}\, \frac{1}{N!}\sum_{(v_1^*,v_2^*)\,\in\,\text{fixed-points}}\, \chi_{(v_1^*,v_2^*)}\,{e^{\mathcal{V}(v_1^*,v_2^*)}}\,.
\end{split}
\ee
The factor~$\chi$ denotes a potential subleading contribution coming from the expansion of the one-loop determinant~$\frac{1}{\mathcal{H}(L)}$ which is inversely proportional to~\cite{Cabo-Bizet:2020ewf}
\be\label{OneLoopDet}
\text{Det}\,\Bigl(\frac{\partial^2 \mathcal{V}}{\partial x_a\partial y_a}\Bigr)\,.
\ee
The~$x_a\equiv v_{1a}+\tau v_{2a}$ and~$y_{a} \equiv v_{2a}$ are two (complex) linear combinations of~$v_{1a}\sim v_{1a}+1$ and~$v_{2a}\sim v_{2a}+1$ that keep the periodicities~$x_a\sim x_a+1$ and~$y_a\sim y_a+1\,$.~\footnote{In this paper we focus only on the Cardy-like limits where the exponential pre-factor~${e^{\mathcal{V}(v^{*}_{1},\,v^{*}_{2})}}$ grows. These Cardy-like limits correspond to the selection of~$M$-wings, in the language of~\cite{Lezcano:2021qbj}. For the limits in the~$W$-wings our conclusions do not apply. For example, in such cases there could be logarithmic corrections (to the effective action) of the form~$\log (m\tau+n)$, as shown in~\cite{Lezcano:2021qbj}, and recently argued in~\cite{Ardehali:2021irq}. In the Cardy-like limits we are studying such corrections are not present~\cite{PaperBulk}.  We thank A. Ardehali for a useful conversation regarding this point.}
The same asymptotic form~\eqref{EquivariantFormula} can be obtained using the Cardy-like expansion of the Bethe ansatz representation~\eqref{BARepIndex}~\cite{Closset:2017bse}\cite{Benini:2018mlo}. Both representations are the same~\cite{Cabo-Bizet:2020nkr}\cite{Cabo-Bizet:2019osg}:\,~\footnote{This is shown in the version 2 of~\cite{Cabo-Bizet:2020nkr}(To appear)} equivariant fixed-points are equivalent to Bethe roots~$L$~\cite{Closset:2017bse}~\cite{Benini:2018mlo}, thus we will use both terms interchangeably.~\footnote{ For example, the configurations~\eqref{SolutionsMNPhases} (below) are the same as the set of Bethe roots originally identified by~\cite{Benini:2018mlo}~\cite{Hong:2018viz,Hosseini:2016cyf}, in the context of the Bethe ansatz formula of~\cite{Benini:2018mlo,Closset:2017bse}.}

The subleading numerical factor~$\chi$ is the same for all the fixed-points that are expected to dominate the Cardy-like limit (which we will introduce below in~\eqref{SolsCardy}), and without loss of generality we can set~$\chi\,=\,1\,$. That all such contributions are the same is a consequence of the double periodicity of~$\mathcal{V}$, and of its second derivatives, independently of the form of~$\mathcal{V}$~\cite{Cabo-Bizet:2019eaf,Cabo-Bizet:2020ewf}.  The~$\mathcal{V}$ can be defined out of a doubly periodic single-particle potential~$\mathcal{V}_0$ as follows
\be
\mathcal{V}\= \sum_{v_1,\,v_2\,=\,1}^{N} \mathcal{V}_0(v_{1a}-v_{1b},\, v_{2a}-v_{2b})\,.
\ee
In this equation~$v_{1N}\=-\sum_{a=1}^{N-1} v_{1a}$ and~$v_{2N}\=-\sum_{a=1}^{N-1} v_{2a}$. The fixed-point conditions are
\be\label{SaddlePoints}
\partial_{v_{1a}} \mathcal{V}\=0\,,\,\quad \partial_{v_{2a}} \mathcal{V}\=0\,,\,\qquad a\=1\,,\ldots\,,\,N-1\,. 
\ee
Independently of the form of~$\mathcal{V}_0$ the set of solutions to~\eqref{SaddlePoints} always include
\be\label{SolutionsMNPhases}
v^{(d,\widehat{q})}_{1a}\= \frac{d a \,+\,\widehat{q}}{N}\,, \,\qquad v^{(c,\widehat{p})}_{2a}\= \frac{c a \,+\,\widehat{p}}{N}\,,\,\qquad a\= 1\,,\ldots\,,\,N\,-\,1\,,
\ee
where~$c$ and~$d$ are integers defined modulo~$N\,$; thus, we can assume them to range in between~$1$ and~$N$. In this paper we assume~$N$ to be prime. The other known solutions group in continuous families~\cite{ArabiArdehali:2019orz} and all indications are that they are exponentially suppressed in the Cardy-like limit~\cite{PaperBulk}. The labels~$\widehat{p}$ and~$\widehat{q}$ are defined modulo~$N$ as well, but they are not always integers, instead
\be
\begin{split}
\widehat{p}\,,\,\widehat{q}&\=0\,,\,\ldots,\, N-1\,,\qquad \text{if~$N$ odd}\,, \\ \widehat{p}\,-\,\frac{1}{2}\,,\,\widehat{q}\,-\,\frac{1}{2}&\=0\,,\,\ldots,\, N-1\,,\, \qquad \text{if~$N$ even}\,.
\end{split}
\ee 
Any smooth double-periodic potential~$\mathcal{V}(x,y)$ (and the one-loop determinant contribution~\eqref{OneLoopDet}) when evaluated on~\eqref{SolutionsMNPhases}, does not depend on the labels~$\widehat{p}$ and~$\widehat{q}$.~\footnote{More precisely, the values of~$\mathcal{V}$ and~\eqref{OneLoopDet} at fixed-points~\eqref{SolutionsMNPhases} are independent of~$\widehat{p}$ and~$\widehat{q}\,$ at fixed~$c$ and~$d$. They do depend non-trivially on the labels~$c$ and~$d\,$. These two previous claims have been checked without using an explicit form for~$\mathcal{V}$, only assuming double periodicity and smoothness in an open domain where the fixed points lie.} Moreover, some solutions in~\eqref{SolutionsMNPhases} are identified under Weyl permutations. For instance, given any~$\ell\in\mathbb{N}$ and~$0\leq\widehat{p}^\star\,,\,\widehat{q}^\star <N$ there exist~$0\leq c,d, p,q<N$ such that
\be\label{Identification0}
v^{(c,d),{(\widehat{p},\widehat{q})}}\,\sim\,v^{( \ell c,\ell d),{(\widehat{p}^\star,\widehat{q}^\star)}}\,.
\ee
Then, the only relevant fixed-points are the
\be
(c,d) \in \mathbb{Z}_N\times \mathbb{Z}_N\,/\,\mathcal{L}
\ee
where~$\mathcal{L}$ is the identification~\eqref{Identification0}, which can be incorporated by requiring $c$ and $d$ to be co-primes, i.e., gcd$(c,d)=1\,$. The set of solutions~\eqref{SolutionsMNPhases} will be called~$(m,n)$ fixed-points for short. 

Using~\eqref{EquivariantFormula}, and the fact that the limit~$\tau\to-\frac{n}{m}$ selects at fixed~$\widehat{p}$ and~$\widehat{q}\,$ those solutions with~$(c,d)=(m,n)\,$, one obtains (after substituting~$\chi\to1$)
\be\label{CardyLimit}
\begin{split}
\mathcal{I}&\,\underset{\tau\,\to\,-\frac{n}{m}}{\Large\simeq}\, \frac{(N-1)!}{N!}\sum_{(\widehat{q}\,,\,\widehat{p})}\,{e^{\mathcal{V}(v^{(n,\widehat{q})}_{1},\,v^{(m,\widehat{p})}_{2})}}\,+\,\text{other fixed-points} \\
&\= N\,{e^{\mathcal{V}(v^{(n,0)}_{1},\,v^{( m,0)}_2)}}\,+\,\text{other fixed-points}\,,
\end{split}
\ee
where the~\emph{Casimir factor}~${e^{\mathcal{V}(v^{(n,\widehat{q})}_{1},\,v^{( m,\widehat{p})}_2)}}$ is by definition the limit~$\tau\to -\frac{n}{m}$ of the integrand of~\eqref{TheIndex}, or equivalently~$\mathcal{F}$, evaluated at the configurations~\eqref{SolutionsMNPhases}\,.~\footnote{The explicit expression for these Casimir factors is $e^{-(N^2-1)\, \mathcal{S}_{(m,n)}(\tau)+\pi\text{i}\mathcal{O}(1)}$ were the function~$\mathcal{S}_{(m,n)}(\tau)$ has been given, for instance, in equation (D.5) of the accompanying paper~\cite{PaperBulk}.}
 
As mentioned before, the potential contributions coming from other fixed-points are conjectured to be suppressed in the Cardy-like limit~\cite{PaperBulk}. These potentially extra contributions vanish for~$m=1$,~$n=0\,$ at any~$N$ (as shown in~\cite{GonzalezLezcano:2020yeb,Amariti:2021ubd,Cassani:2021fyv,ArabiArdehali:2021nsx}, see also the discussion in the accompanying paper~\cite{PaperBulk}). They also vanish when~$N=2$~\cite{Benini:2021ano}\cite{Lezcano:2021qbj}. Although we have not yet checked it for generic~$m$ and~$n\,$, based on constraint~\eqref{DivisorConstraint} and on the fact that only~$N$ out of the subset of fixed points~\eqref{SolutionsMNPhases} dominates in the limit~$\tau\to-\frac{n}{m}$, we expect that even if the \emph{other fixed points} contribute at leading order in the expansion~$\tau\to -\frac{n}{m}$, the gauge orbits associated to the~$N$ dominating ones within the family~\eqref{SolutionsMNPhases} forms a representation of the Verlinde algebra by themselves (in the way explained in section~\ref{sec:QuantumGroup}).

Given this expectation, from now on, we will focus on the contributions coming from the fixed-points~\eqref{SolutionsMNPhases}. The~$N$ dominating solutions within~\eqref{SolutionsMNPhases} in the expansion~$\tau\,\to\,-\,\frac{n}{m}\,$ are counted by the overall factor of~$N$ in the second line of~\eqref{CardyLimit}. The latter factor comes from the fact that at fixed~$c=m$ and~$d=n\,$, there are
\be\label{PreFactor}
N^2\times (N-1)!
\ee
solutions, all of them with the same potential~$\mathcal{V}$. The~$N^2$ comes from the~$N$ possible values of~$\widehat{p}$ times the~$N$ possible values of~$\widehat{q}\,$. The~$(N-1)!$ comes from the permutations of the label~$a$. The remnant factor of~$N$ comes from dividing~\eqref{PreFactor} by~$N!$. 

The remaining~$N$ solutions can be parametrized by the following choice of~
\be\label{ChoiceDiagonal}
{\widehat{p}}\= m \,\widehat{a} \,\text{mod}\,N\,, \qquad{\widehat{q}}\= n\, \widehat{a} \,\text{mod}\,N\,,\qquad \widehat{a}\=0\,,\,1\,,\,\ldots\,,\, N-1\,.
\ee
Note that the index~$\widehat{a}$ is independent of~$a\,$. Other parametrizations are possible but we find convenient to use~\eqref{ChoiceDiagonal}.

\eqref{ChoiceDiagonal} corresponds to the following~$N$ dominant solutions (in Cardy-like limit~$\tau\to-\frac{n}{m}$)
\be\label{SolsCardy}
v_a\= \frac{(a-\widehat{a})\,\text{mod}\,N}{N}\, (m \tau\,+\,n)\,,
\ee
where for each of the solutions, labelled by~$\widehat{a}\,=\,0,\ldots N-1\,$, the index~$(a\,-\,\widehat{a})\, \text{mod}\,N$ ranges over~$N-1$ different integer values corresponding to~$a=1,\ldots N-1\,$. The different solutions correspond to the possible sets of~$N-1$ different integers in between~$0$ and~$N-1$ (without minding the ordering). For instance, for~$N=3$ the unique such~$3$ solutions, labelled by the index~$\widehat{a}=0,1,2$, respectively, are
\be
\{\{v_a\}\}\=\{\{0\,,\, \frac{1}{3}(m\tau+n)\}\,,\,\{\frac{1}{3}(m\tau+n)\,,\, \frac{2}{3}(m\tau+n)\}\,,\,\{0\,,\,\frac{2}{3}(m\tau+n)\}\}\,.
\ee

\section{A quantum phase is defined by~$N$ condensates}
\label{sec3}

This section shows that the fixed-points~\eqref{SolsCardy} can be understood as bound states of two Dyonic surface condensates located at the north and south fixed loci of the rotational~$U(1)$ action. These condensates carry a specific electromagnetic $\mathbb{Z}^{(1)}_N$ one-form charges.

\subsection*{Center and 1-form symmetries} \label{OneFormSymmetries}

This subsection briefly introduces the concept of center (and one-form) symmetries.

Focus on gauge potentials~$\widetilde{A}_\mu(x)$ for which there exists a point in their gauge orbit where only the Cartan projection~$A_\mu(x)\,$ is non vanishing. We will call these~\emph{Abelian orbits}, and the point at which only Cartan elements are non-vanishing, the~\emph{Abelian ansatz}. A generic point in a Abelian orbit~$\widetilde{A}_\mu(x)$ can be recovered \emph{via} regular gauge transformations i.e.
\be\label{Orbits}
\begin{split}
\widetilde{A}_\mu(x)&\= G^{-1}\, A_\mu(x)\,G \,-\,\i G^{-1}\partial_{\mu} \, G\,. \\
A_\mu(x)&\,\equiv\,A_{a\mu}(x) T^a\,.
\end{split}
\ee
The~$T_a$ are elements in a Cartan subalgebra of~$SU(N)$. For the matter content of~$\mathcal{N}=4$ SYM~$T_a$'s are in the adjoint representation, but as we mentioned before there can be emergent or probe extended operators in other representations~$\textbf{r}$.

It is convenient to define the holonomy integrals along the time cycle
\be\label{PolyakovLineIntegrals}
v_a\=\oint_{t_E} A_{a\mu}\, dx^\mu\,, \qquad a\=1\,,\,\ldots\,,\,\text{rank}G\,.
\ee
As already mentioned, these are not gauge invariant variables but they carry gauge invariant information about the gauge orbits. Take a generic point~$\widetilde{A}$ in the orbit. The path ordered exponential (which is closely related to the monodromy operators at the punctures~$M_{\ell}$)
\be
\Pi\=P e^{\oint_{t_E} \widetilde{A}_{\mu}\, dx^\mu}
\ee
transforms covariantly under regular gauge transformations. Regular gauge transformations carry trivial holonomy along any cycle, thus~$\underset{\textbf{r}}{\text{Tr}}\Pi$, for any~$\textbf{r}$, is an invariant of the gauge orbit.

Under a singular gauge transformation~$G\,$,
\be
\Pi \to G_f^{-1} \Pi G_i\,,
\ee
where~$G_i$ is the value of~$G$ at a starting point of the cycle and~$G_f$ is the value of~$G$ at the ending point of the cycle.~If~$G$ is such that~$G_i G_f^{-1} \= e^{\i \Phi_{G,\textbf{r}}}$
belongs to the center of~$G$, then, 
\be
\underset{\textbf{r}}{\text{Tr}} \Pi \underset{G}{\longrightarrow} e^{\i\Phi_{G,\textbf{r}}}\,\underset{\textbf{r}}{\text{Tr}} \Pi\,,
\ee
and~$G$ is said to generate a center transformation of the gauge group. For~$SU(N)$ the center group is~$\mathbb{Z}_N\,$: then a center transformation~$G$ on a Polyakov loop can be represented as follows
\be\label{CenterSym}
\underset{\textbf{r}}{\text{Tr}} \Pi \underset{G}{\longrightarrow} e^{\frac{2\pi\i m }{N}\,{n}_{\textbf{r}}}\,\underset{\textbf{r}}{\text{Tr}} \Pi\,,
\ee
where the integer~$m \sim m +N \in \mathbb{Z}$ depends on~$G$, but not on the representation~$\textbf{r}$. The integer~$n_\textbf{r}\sim n_\textbf{r} + N$ depends on~$\textbf{r}\,$, it is its~$N$-ality.~For the fundamental representation one can define~$n_{\textbf{r}}\=1$. For the anti-fundamental representation, which is the complex conjugated of the fundamental,~$n_\textbf{r}\=-1\,$. For the adjoint, which enters in the product of fundamental and anti-fundamental, due to the fact that the overall phase of a tensor product representation is the product of the overall phases of each factor representation,~$e^{2\pi\i (m-m)}\=1$,~$n_{\textbf{r}}\=0$. For a generic representation~$\textbf{r}$, the~$n_\textbf{r}$ happens to be equal to the number of fundamental representations that are needed to construct~$\textbf{r}$ (modulo~$N$); for example, for the anti-fundamental such number is~$-1+N$, and for the adjoint,~$0+N$. 

Although the adjoint holonomy operator does not detect changes under center transformations, that does not mean that in a theory such as~$\mathcal{N}=4$ SYM, center symmetry can not be broken, it means that one can not use an adjoint holonomy operator to detect the breaking. To detect center symmetry breaking for adjoint matter content, one can insert a probe operator, such as the \emph{Polyakov loop},~\footnote{A Polyakov loop is a Wilson loop extended along the~$t_E$-time cycle.} 
\begin{equation}
\frac{1}{\text{dim}\textbf{r}}\,\underset{\textbf{r}}{\text{Tr}} \Pi\,,
\end{equation}
with, say,~$\textbf{r}$ being the fundamental of~$SU(N)\,$. If the expectation value of the probe operator vanishes, then center symmetry is preserved, as transformation~\eqref{CenterSym} does nothing on zero. On the contrary, if for any single one of such operators, the expectation value is different from zero, center symmetry is broken. For example, insert a probe Polyakov loop in representation~$\textbf{r}$. Evaluated at the gauge orbit~\eqref{Orbits}, the latter is only a function of the Cartan ansatz~\eqref{Cartan}
\begin{equation}
\frac{1}{\text{dim}R}\,\underset{\textbf{r}}{\text{Tr}} \Pi\,=\, \chi_{\textbf{r}}( \i v)\=\frac{1}{\text{dim}\textbf{r}}\,\sum_{\rho \,\in\, \textbf{r}}\,e^{\i \rho(v)}\,.
\end{equation}
~$\chi_\textbf{r}$ is the character of the representation~$\textbf{r}$ of~$SU(N)$, and the~$\rho$'s are the weight vectors of~$\textbf{r}\,$. For generic~$v\,$, the center symmetry is broken. But sometimes it is not. Assume~$v$ localizes to the~$(m,n)=(0,1)$ fixed-point.  If one assumes~$\textbf{r}$ to be the fundamental representation, the vacuum expectation value of the Polyakov loop vanishes. As every other representation is a tensor product of the fundamentals, the result remains the same for any other representation. Thus, center symmetry is unbroken in the~$(0,1)$ phase.

We have only mentioned the center symmetry associated to the Polyakov cycle. More generally, one can define a Wilson loop operator along other one-cycles, and draw similar conclusions as for the Polyakov loop. In the formulation of~\cite{Gaiotto:2014kfa}, the associated center symmetries are called~$\mathbb{Z}_N^{(1)}$ discrete one-form symmetries.

\subsection{Complex fixed-points~$\,=\,$ Singular flat potentials}\label{CondensatesCharges}

Take the tangential differential to the family of twisted time cycles~$\Gamma$
\begin{equation}\label{TheBase}
\diff x^\mu(\a)\,\equiv\,(\diff t_E,\diff \theta,\diff \phi_1,\diff \phi_2)\=\bigl(\diff\a,0,\,-\frac{2 \pi \tau}{\beta}\, \diff \a,-\frac{2 \pi \tau}{\beta}\, \diff \a \bigr)\,.
\end{equation}
Define the holonomy integral~$\int_{\Gamma} A_\mu\, dx^\mu_\Gamma \,\equiv\,\int_{0}^{\beta} \diff \a\, A_\mu (x_\Gamma(\alpha)) \frac{dx^\mu_\Gamma(
\a)}{\diff \a} \,$. Consider the simple Abelian ansatz
\be\label{SimplleAnsatz}
(A_{a})_{t_E}\= \text{const}\,,\,(A_{a})_{\phi_{1}}\=\text{const}\,,\, (A_{a})_{\phi_2}\=\text{const} \,.
\ee~\footnote{Flat gauge potentials require~
\be\label{RegCond}
(A_{a})_{\theta}\,=\,(A_{a})_{\phi_1}\,=\,(A_{a})_{\phi_2}\,=\,0\,.
\ee
} 
At the moment~$(A_{a})_{\theta}$ is not constrained, but eventually, it will be fixed to zero.~\footnote{It will be fixed to zero by the BPS condition in the punctured geometry. } 

The index~\eqref{N4SYM} is obtained after regularizing a quotient of infinite products of fermionic and bosonic eigenvalues of the twisted Hamiltonian~\cite{Cabo-Bizet:2018ehj}
\be
{\widetilde{H}}\,\equiv\, \frac{\diff x^\mu_{\Gamma}(\a)}{\diff a} \,\bigl(-\i\,\nabla_{\mu}\bigr)\=\nabla_{t_E}\,-\,\frac{2\pi\tau}{\beta}\,\nabla_{\phi_1}\,-\,\frac{2\pi\tau}{\beta}\, \nabla_{\phi_2}\,.
\ee
~${\widetilde{H}}$ generates translations along the twisted time cycle. It is also (up to a c-number normalization) the square of the supercharge used to localize the original supersymmetric path integral. In this expression~$\nabla$ represents the fully covariant derivative. From how the gauge connection enters in the covariant derivatives~$\nabla_\mu\=\ldots \,+\,\i A_{\mu}\,+\,\ldots\,$, it follows that the supersymmetric partition function (the index) depends only on a specific combination of~$A_{a\mu}\,$: for a constant connection, as we will assume, such combination is
\be\label{Identification}
{ v_a \= v_{a1}\,+\,\tau v_{a2}\,\equiv\, \int_{\Gamma} (A_{a})_{\mu}\, dx^\mu}\,,
\ee
with the periodic identifications~$v_{a 1}\sim v_{a 1}+1\in\mathbb{R}$ and~$v_{a_2}\sim  v_{a 2}+1\in\mathbb{R}\,$. 
From~\eqref{TheBase} and~\eqref{Identification} it follows that
\be\label{AphiVEV}
\begin{split}
v_{a1}&\=\int_{0}^\beta\diff t_E\,(A_{a})_{t_E}\,, \qquad v_{a2}\,=\,-\,\int_0^{2\pi}\diff\phi_1 (A_{a})_{\phi_1}\,-\,\int^{2\pi}_{0} \diff \phi_2 (A_{a})_{\phi_2}\,.
\end{split}
\ee
These identifications imply that
\be\label{ComplexSingular}
v_{a2}\,\neq\,0\,\implies\, (A_{a})_{\phi_1}+(A_{a})_{\phi_2}\,\neq\,0\,.
\ee
Assuming an ansatz for~$A_{a\mu}$ invariant under the transformation~$\bigl(\theta,\phi_1\bigr)\, \to\, \bigl(\frac{\pi}{2}-\theta,\,\phi_2\bigr)\,$,
which is a discrete isometry of~\eqref{background_metric0} for~$\tau\,=\,\sigma\,$, one obtains
\be\label{ComplexV2Implication}
\int_0^{2\pi}\diff\phi_1 (A_{a})_{\phi_1}\=\int_0^{2\pi}\diff\phi_2 (A_{a})_{\phi_2}\=-\,\frac{1}{2}\,{v_{a2}}\,.
\ee
~\eqref{ComplexSingular} and~\eqref{ComplexV2Implication} say that a fixed-point with~$v_{2a}\,\neq\,0$, a complex fixed-point, can be interpreted as a singular gauge configuration in spacetime~\eqref{background_metric00}.

\subsection{Magnetic and Electric cycles: fluxes}\label{subsec:MagElect}

This subsection completes the interpretation of the fixed-points~\eqref{SolsCardy} as bound-states of two surface operators localized at two punctures~$\theta=0$ and~$\theta=\frac{\pi}{2}$. At the level of the geometry we will fix~$\tau=\sigma=-\frac{n}{m}+$ and infinitessimal number with~$m$ and~$n$ coprimes. That is the set up at which fixed-points~\eqref{SolsCardy} dominate the BPS limit of the thermal partition function.

\paragraph{Defining cycles and relations} To measure magnetic flux, we consider two families of cycles, denoted as~$\Gamma_{m \text{N}}$ and~$\Gamma_{m \text{S}}$
\begin{equation}\label{phicycle0}
\begin{split}
x^\mu_{\Gamma_{m \text{N}}}(\a)&\,\equiv\,\bigl(\theta\,=\, g(\a)\,,\, \phi_1\=\a\,\bigr)\,, \\
x^\mu_{\Gamma_{m \text{S}}}(\a)&\,\equiv\,\bigl(\theta\,=\, \frac{\pi}{2}-g(\a)\,,\, \phi_2\=\a\bigr)\,,
\end{split}
\end{equation}
with worldline parameter~$0\,\leq\,\a\,<\,2\pi\,$. We have omitted the components that remain fixed as~$\alpha$ varies.~$g$ is a real function that defines both, distance from the \emph{poles} ($\theta=0$ and~$\theta=-\frac{\pi}{2}$) and wiggling.
The label~$m$ means~\emph{magnetic} and the~$N$ (resp. $S$) means \emph{north} (resp. \emph{south}). Both cycles~$\Gamma_{m\text{N}}$ and~$\Gamma_{m\text{S}}$ are boundary of two-dimensional regions
\be\label{BoundaryMagnetic}
\Gamma_{m\text{N}}\=\partial \Sigma_{\Gamma_{m\text{N}}}\,,\, \qquad \Gamma_{m\text{S}}\=\partial \Sigma_{\Gamma_{m\text{S}}}\,,
\ee
with
\begin{equation}
\begin{split}
\Sigma_{\Gamma_{m\text{N}}}&\=\Bigl( 0\,\leq\,\theta\,\leq\, g(\a)\,,\, \phi_1\=\a\Bigr)\,, \\
\Sigma_{\Gamma_{m\text{S}}}&\=\Bigl(\,\frac{\pi}{2}\,-\,g(\a)\,\leq\,\theta\,\leq\,\frac{\pi}{2}\,,\, \phi_2\=\a\,\Bigr)\,.
\end{split}
\end{equation}
~\footnote{Coordinates components normal to the surface are not shown in these expressions.} These two regions are centered at the fixed loci of the rotation.

To measure electric flux we use two families of cycles,~$\Gamma_{e\text{N}}$ and~$\Gamma_{e\text{S}}\,,$
\begin{equation}\label{tAlternativecycle0}
\begin{split}
x^\mu_{\Gamma_{e\text{N}}}(\a)\,&\equiv\,\bigl( \,t_{E}\,\= \a\,,\, \theta\=g(\a)\,,\, \phi_2\=-\,\frac{2\pi \tau}{\beta} \alpha\,\bigr)\,, \\
x^\mu_{\Gamma_{e\text{S}}}(\a)\,&\equiv\,\bigl( \,t_{E}\,\= \a\,,\, \theta\=\frac{\pi}{2}\,-\,g(\a)\,,\, \phi_1\=-\,\frac{2\pi \tau}{\beta} \alpha\bigr)\,,
\end{split}
\end{equation}
with worldline parameter~$0\,\leq\,\a\,<\,m\,\times\, \beta\,$. These cycles wind~$m$ times over the base~$S_1$ (the twisted time cycle) and~$n$ times over a cycle on the base~$S_3$ around~$\theta=0$ (resp. $\theta=\frac{\pi}{2}$). This is illustrated in figure~\ref{Loops}. 
\begin{figure}[h]\centering
\includegraphics[width=10cm]{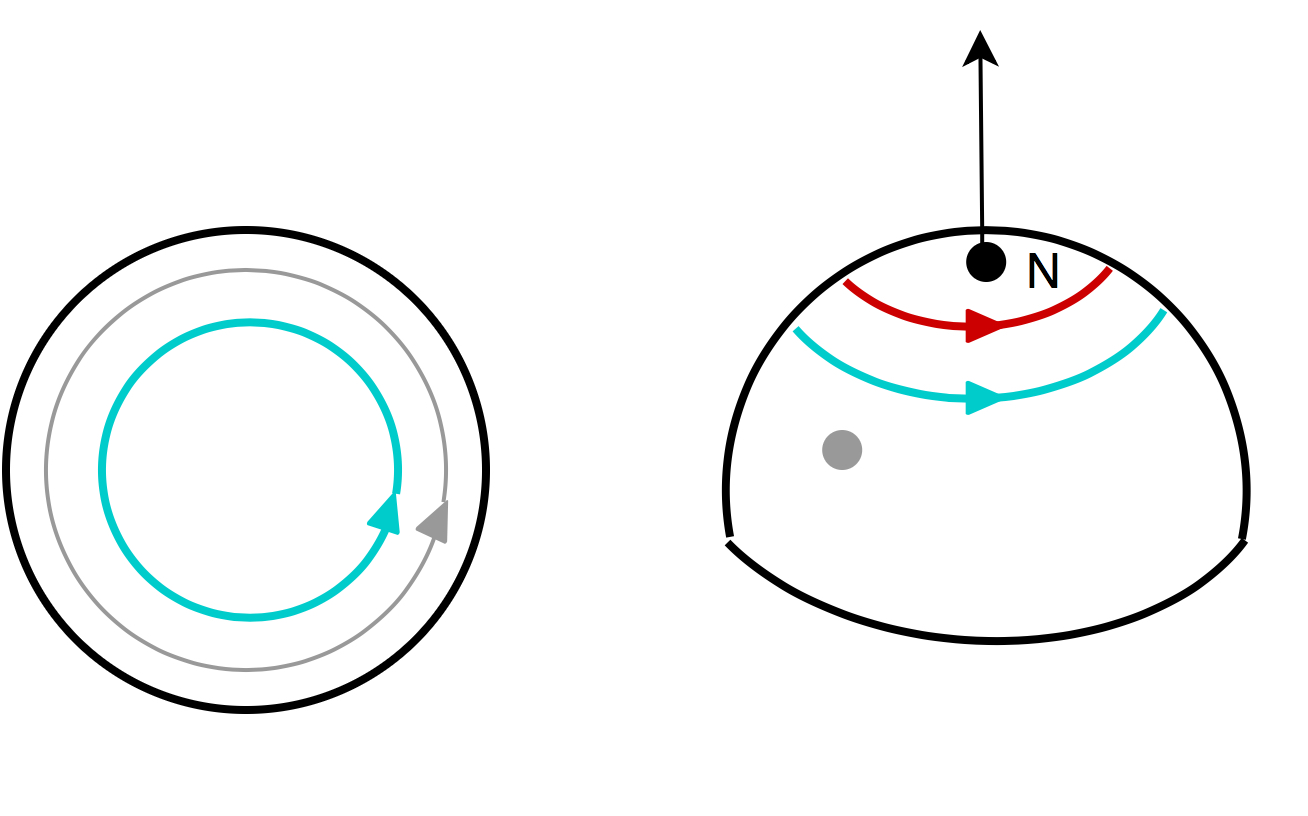} 
\caption{Schematic representation of the twisted time (grey), electric north (blue) and magnetic north (red) cycles in the singular coordinates where the fibration looks trivial. This representation is obtained after implementing the singular change of coordinates~$\phi_{1,2}\to\phi_{1,2}-\frac{2\pi \tau}{\beta}t_E$ (to the \emph{rest frame}). The loops in the hemisphere represent two-tori. One of the cycles of the tori contracts at the north pole; the other one remains finite size. The magnetic and electric cycles move along the former and are static on the latter. The twisted time cycle does not move along the~$S_3$ in the rest frame. The figure does not represent the electric cycle's non-trivial winding numbers, which are~$m$ around the base~$S_1$ and~$n$ around the north locus represented with the letter~$N$ in the figure. When located at the north (resp. south) locus~$N$ (resp.~$S$) the electric north (resp. south) cycle is identical to~$m$ copies of the twisted time cycle.  }
\label{Loops}
\end{figure}
The electric and twisted time cycles obey the following relation
\be\label{ElectricRelations1}
\begin{split}
\Gamma_{e\text{N}}&\,-\, m \times \Gamma\=0 \qquad \text{at}\,\qquad \theta\=0\,, \\
\Gamma_{e\text{S}}&\,-\, m \times \Gamma\=0 \qquad \text{at}\,\qquad \theta\=\frac{\pi}{2}\,.
\end{split}
\ee
These relations hold because the difference between~$\Gamma$ and~$\Gamma_{e \text{N}}$ (resp.~$\Gamma_{e \text{S}}$) comes from their relative motion along the~$\phi_1$ (resp.~$\phi_2$) cycle, as well as from the fact that the latter cycle contracts at~$\theta=0$ (resp.~$\theta=\frac{\pi}{2}$). Instead, for~$\theta\neq0\,$ (resp. $\theta\neq\frac{\pi}{2}\,$), depending on~$m$ and~$n$, there are two possibilities, either:
\be\label{orientable}
\Gamma_{e\text{N}} \,-\,m\times \Gamma_{\theta=0}\= \partial \Sigma_{\Gamma_{eN}}\,,\qquad \Gamma_{e\text{S}} \,-\,m\times \Gamma_{\theta=\frac{\pi}{2}}\= \partial \Sigma_{\Gamma_{eS}}\,,
\ee
or
\be\label{unorientable}
\Gamma_{e\text{N}} \= \partial \Sigma_{\Gamma_{eN}}\,,\qquad \Gamma_{e\text{S}} \= \partial \Sigma_{\Gamma_{eS}}\,.
\ee
In these expressions
\be
\begin{split}
\Sigma_{\Gamma_{e\text{N}}}&\=\Bigl( \,t_{E}\= \a\,,\, 0\,\leq\,\theta\,<\,g(\a)\,,\, \phi_2\=-\,\frac{2\pi \tau}{\beta} \alpha\,\Bigr)\,, \\
\Sigma_{\Gamma_{e\text{S}}}&\=\Bigl(\,t_{E}\= \a\,,\, \frac{\pi}{2}\,-\,g(\a)\,\leq\,\theta\,<\,\frac{\pi}{2}\,,\, \phi_1\=-\,\frac{2\pi \tau}{\beta} \alpha\,\Bigr)\,.
\end{split}
\ee
\begin{figure}[h]\centering
\includegraphics[width=6.8cm]{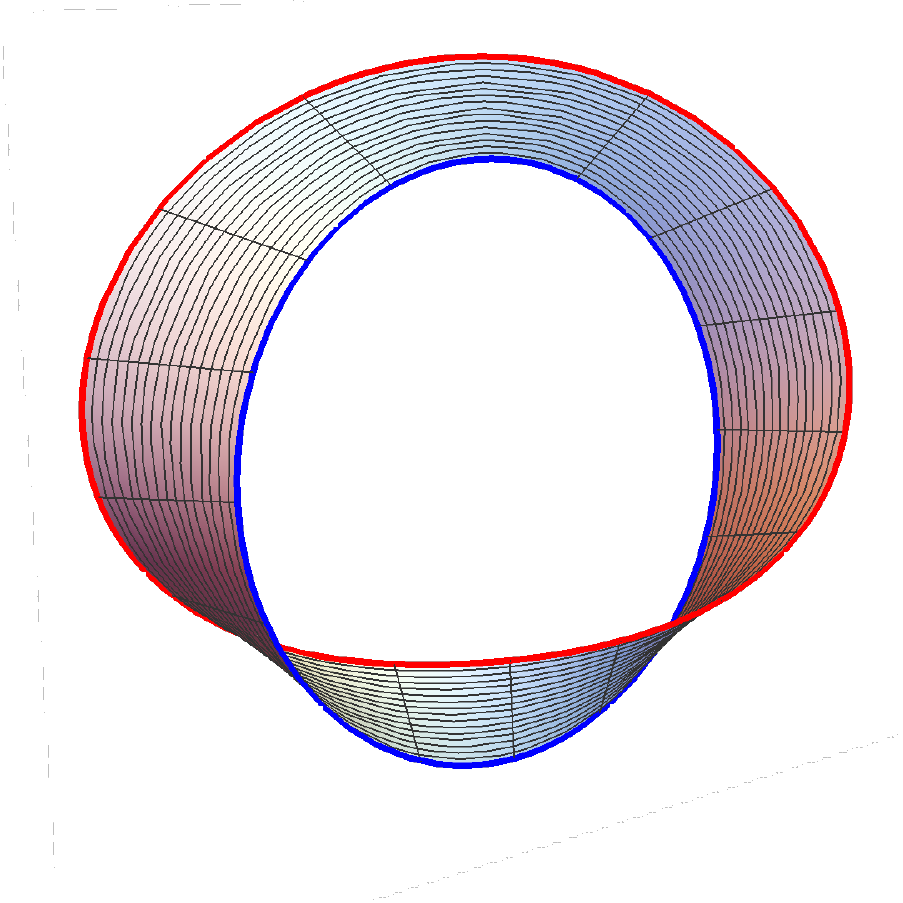} 
\quad \includegraphics[width=7.5cm]{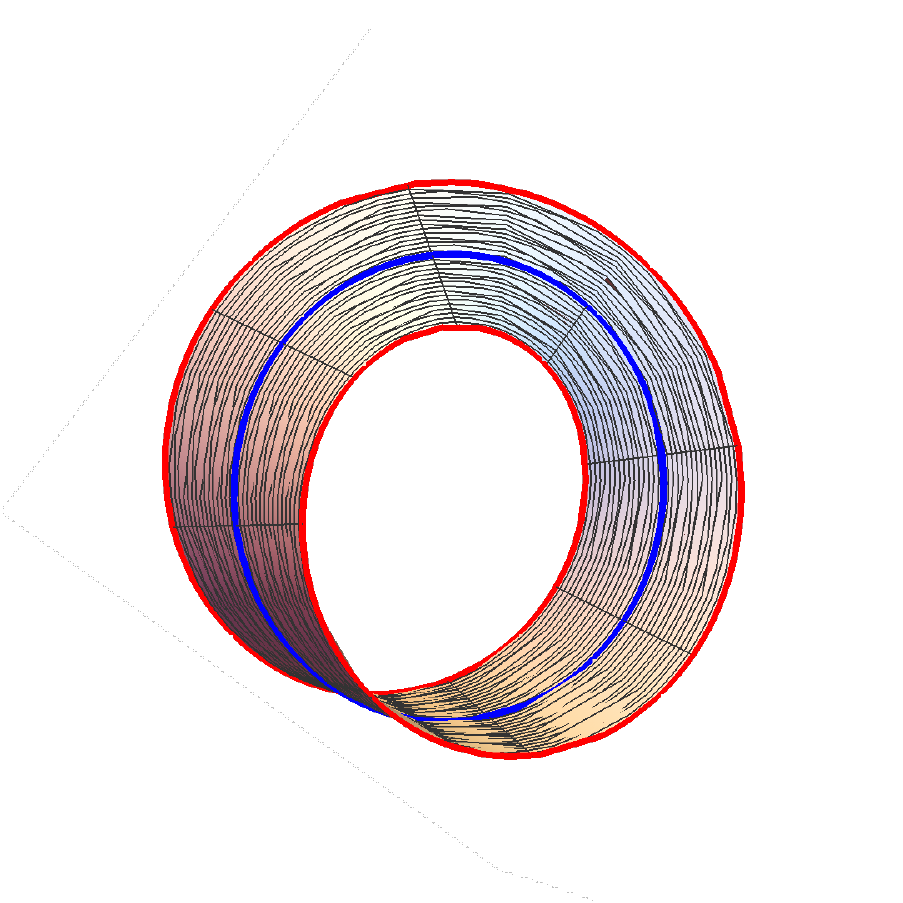} 
\caption{ The blue cycle represents the twisted time cycle~$\Gamma$ located at~$\theta=0$ (resp.~$\theta=\pi$)~$\Gamma_{\theta=0}$ (resp.~$\Gamma_{\theta=\frac{\pi}{2}}$). The red cycle represents the electric cycles~$\Gamma_{eN}$ (resp. $\Gamma_{eS}$).  The ribbons represent the regions~$\Sigma_{eN}$ (resp. $\Sigma_{eS}$). The figure to the left corresponds to the case~$(m=1,n=2)$. In this case the ribbon is orientable and its boundary is~$\Gamma_e -m \Gamma_{\theta=0}$. The figure to the right corresponds to the case~$(m=1,n=1)$. In this case the ribbon is unorientable and its boundary is~$\Gamma_{eN}$~($\Gamma_{eS}$). The case~$(m=1, n=0)$ corresponds to un-knotting the cycles in the figure to the left. }
\label{fig:Topol}
\end{figure}
Relations~\eqref{orientable} correspond to values of~$m$ and~$n$ for which the regions~$\Sigma_{eN}$ and~$\Sigma_{eS}$ are orientable surfaces (Left  figure~\ref{fig:Topol}). Relations~\eqref{unorientable} correspond to the values of~$m$ and~$n$ for which~$\Sigma_{eN}$ and~$\Sigma_{eS}$ are unorientable surfaces (Right figure~\ref{fig:Topol}).

\subsubsection*{Chromomagnetic flux}
Let us measure the fluxes.
The differential of the magnetic cycle~\eqref{phicycle0} is of the form
\begin{equation}\label{phicycle}
\diff x^\mu_{\Gamma_{m \text{N}}}(\a)\,\equiv\,(\diff t_E,\diff \theta,\diff \phi_1,\diff \phi_2)\=\bigl(0,\diff g(\a),\, \diff(\a),\, 0 \bigr)
\end{equation}
where~$g$ is a real and smooth function such that~$0<g(0)=g(2\pi)<\frac{\pi}{2}\,$, and~$\a\in[0, 2\pi)\,$. If we assume no wiggling i.e.~$\frac{\diff g}{\diff \alpha}\,\equiv\,0$
\be
\oint_{{\Gamma_{m \text{N}}}}\, (A_a)_\mu \diff x^\mu \,= \, \int_0^{2\pi}\diff\phi_1 (A_a)_{\phi_1}\,= \, -\,\frac{1}{2}\, v_{a2}\,.
\ee
On the other hand in case~$b)$\,,\,$\oint_{\Gamma_{m \text{N}}}\, A_\mu \diff x^\mu \,=\, 0\,$.
Equation~\eqref{BoundaryMagnetic} and Stokes theorem imply
\begin{align}\label{flux}
\int_{\Sigma_{mN}} \, (F_a)_{\theta \phi_1}\= -\,\frac{1}{2}\,v_{a2}\,.
\end{align}
Equation~\eqref{flux}, which holds for any size of~$\Sigma_{m\text{N}}$ as long as its boundary~$\partial \Sigma_{m\text{N}}$ does not wiggle in the~$\theta$-direction, implies that for every~$\Sigma_{m\text{N}}$, even for those with wiggling boundary~$\partial \Sigma_{m\text{N}}\,$,
\be\label{flux2}
\int_{\Sigma_{mN}} \, (F_a)_{\theta \phi_1} \,\diff \theta\wedge \diff\phi_1 \,\limNorth\,-\,\frac{1}{2}\,v_{a2}\,.
\ee
Had~$\Gamma_{mN}$ not surrounded the north cycle
\be
\int_{\Sigma_{mN}} \, (F_a)_{\theta \phi_1}\,\to\,0\,.
\ee
in the limit in which~$\Gamma_{mN}$ contracts to a point.~\footnote{This is because for bounded gauge potentials the holonomy integral along any such infinitesimal loop is proportional to the difference in between the values of the angular coordinate~$\phi_1$ of the initial and final point of the loop, if such coordinates are the same, as it is the case for a loop not surrounding the north locus, then such a loop integral vanishes in the zero-area limit.} These observations imply that
\be\label{FthhetaPHi}
(F_{a})_{\theta \phi_1} \, = \,-\, \frac{m}{2}\,\frac{ a-\widehat{a}}{N}\,\frac{1}{2\pi}\, \delta(N)\,+\,\text{integrable contribution}\,,
\ee
where~$\delta(N)\,\equiv\, \delta(\theta)$ and~$\delta$ is a Dirac delta, normalized as follows~$\int_{0}
^{\theta} \,\diff \theta^\prime\, \delta(\theta^\prime)\= 1\,$.
We should also note that the (localized) magnetic field lines associated to the configurations~\eqref{FthhetaPHi} are directed along a closed cycle, thus there is no isolated magnetic charge localized at the north locus, just flux. 

The previous analysis can be extended to the south locus and one obtains, when there is no wiggling,
\be
\int_{\Gamma_{m\text{S}}}\, (A_a)_{\mu} \diff x^\mu \= \int_0^{2\pi}\diff\phi_2 (A_a)_{\phi_2}\,= \, -\,\frac{1}{2}\, v_{a2}\,.
\ee
Following analogous steps as before, one obtains
\be\label{ConditionMagSouth}
(F_{a})_{\theta \phi_2} \= -\, \frac{1}{2} \frac{m (a- \widehat{a})}{N}\,\frac{1}{2\pi}\,\delta(S)\,+\,\text{integrable contribution}
\ee
where~$\delta(S)\,\equiv\, \delta(\theta-\frac{\pi}{2})$ and~$\delta$ is the laterally normalized Dirac delta function defined above but this time centered at south.

~\eqref{FthhetaPHi} and~\eqref{ConditionMagSouth} define (magnetic) singular boundary conditions for the gauge potentials at the north and south fixed loci of the rotation.

Requiring the Cartan ansatz to satisfy the BPS conditions in the punctured spacetime, with all scalar VEVs fixed to zero, forces the gauge potentials to be locally flat away from the fixed loci of the rotation.~\footnote{ Following the conventions of~\cite{Nawata:2011un}, and for the spacetime metric used there, which relates to~\eqref{background_metric0} by a local diffeomorphism transformation, the Euclidean~$\frac{1}{16}$-BPS conditions take the form: \begin{equation}\label{BPSConditions}
\begin{split}
F_{03}\,+\,F_{12}&\,=\, -\frac{1}{2}\Bigl[\phi^j,\,\overline{\phi}_j\Bigr]\,, \quad F_{02}\,+\,F_{31}\,=\, F_{01}\,+\,F_{23}\,\,=\,0\,,\\
(\i D_1\,-\,D_2) \,\overline{\phi}_j&\,=\,0\,\quad (D_0\,+\,\text{i}\,D_3\,+\,1)\,\overline{\phi}_j\,=\,0\,, \quad  [\overline{\phi}_i,\,\overline{\phi}_j]\,=\,0\,.
\end{split}
\end{equation}
Thus, for vanishing scalars the gauge potential must be flat for a classical configuration to be supersymmetric.
} That condition fixes the integrable contribution in~\eqref{FthhetaPHi} and~\eqref{ConditionMagSouth} to zero. 
Thus the magnetic flux measured by a fundamental probe Wilson line is independent of the area enclosed by the line, as long as the latter encloses a puncture, otherwise the measured flux is zero (in strict Cardy-like limit).

\subsubsection*{Chromoelectric field strength}

\begin{figure}[h]\centering
\includegraphics[width=9cm]{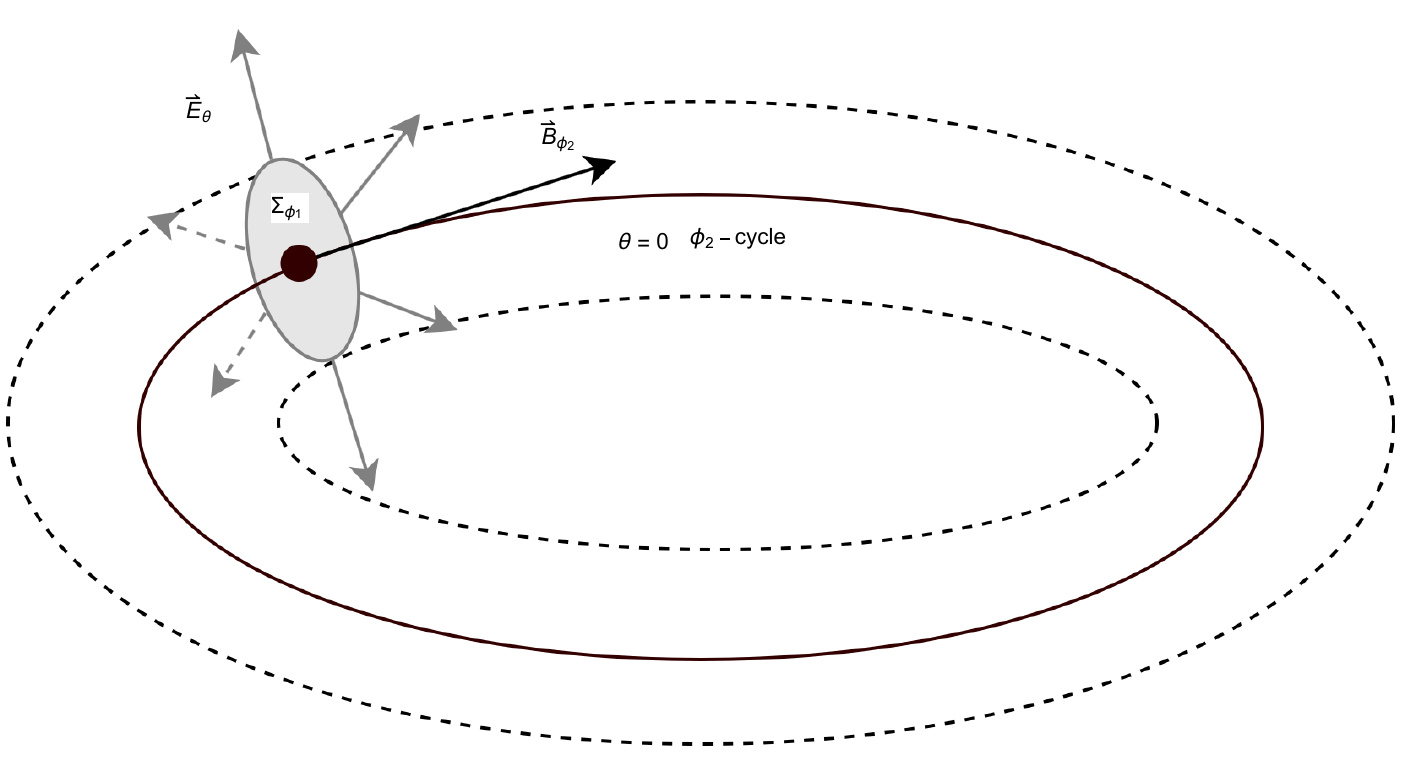} 
\caption{ The bound-state is made of a pair of two-dimensional components carrying Dyonic flux localized at the fixed-point of the rotational~$U(1)$ action. They arise due, in part, to the rotation of~\eqref{background_metric00}. Flatness condition away from the fixed loci of the rotation implies the absence of isolated electric and magnetic charge. The time cycle is not shown in this figure.}
\label{fig:SolidTorus}
\end{figure}
Let us measure the electric flux. The tangential differential to the cycle~$\Gamma_{e \text{N}}$ in~\eqref{tAlternativecycle0} is
\begin{equation}\label{tAlternativecycle}
dx^\mu_{\Gamma_{e\text{N}}}(\alpha)\,\equiv\,(\diff t_E,\diff \theta,\diff \phi_1,\diff \phi_2)\=\bigl(\diff f(\a),\diff g(\a),\, 0,\, -\,\frac{2\pi \tau}{\beta} \diff \a\, \bigr)
\end{equation}
where~$g$ is a real and smooth function such that~$0<g(0)=g(m\beta)<\frac{\pi}{2}\,$, and~$\a\in[0, m\beta)\,$. 

When~$\Gamma_{eN}$ does not wiggle
\be\label{GammaNorthA}
\begin{split}
\int_{\Gamma_{eN}} (A_a)_\mu dx^\mu&\=\,\int^{m \beta}_0 d\alpha \,\Bigl( {(A_a)}_{t_E} \frac{\diff t_E}{\diff \alpha} \,+\,  (A_a)_{\phi_2} \frac{\diff \phi_2}{\diff \alpha} \Bigr), \\
&\=\, m\,\Bigl({v_{1a}}\,+\,
{\tau}\,\frac{v_{a2}}{2}\Bigr)\,\=\, m\,\Bigl({v_{1a}}\,-\,
\frac{n}{m}\,\frac{v_{2a}}{2}\Bigr)\,.
\end{split}
\ee
The second line comes from using 
~\eqref{AphiVEV},~\eqref{ComplexV2Implication},~\eqref{tAlternativecycle}, and~$\tau=-\frac{n}{m}\,+\,$ an infinitesimal number. The effect of wiggling is irrelevant in the zero area limit~$g\to0^+$. Then~\eqref{GammaNorthA} implies that for any~$\Gamma_{e\text{N}}$
\be\label{LimGNA}
\int_{\Gamma_{eN}} (A_a)_\mu dx^\mu \,\limNorth\,  m\,\Bigl({v_{1a}}\,-\,
\frac{n}{m}\,\frac{v_{2a}}{2}\Bigr)\,.
\ee
Assume the choice of~$m$ and~$n$ corresponds to an unorientable~$\Sigma_{e\text{N}}$. Then, using relation~\eqref{unorientable} and Stokes theorem one obtains
\be\label{StokesGN}
\begin{split}
\int_{\Gamma_{e\text{N}}} \,(A_a)_\mu\, \diff x^\mu &\= \int_{\Sigma_{\Gamma_{e\text{N}}}}\, (F_a)_{\theta t_E}\, \diff\theta\,\wedge\, \diff t_E  \,.
\end{split}
\ee
From~\eqref{LimGNA} and~\eqref{StokesGN} it follows that
\be\label{Electric1}
\int_{\Sigma_{\Gamma_{e\text{N}}}}\, (F_a)_{\theta t_E}\,\diff\theta\,\wedge\, \diff t_E  \,\limNorth\,\Bigl(v_{1a}\,-\,\frac{n}{2m}\, v_{2a}\Bigr)\,.
\ee
Moreover, for the same reason mentioned above for the magnetic cycles~$\Gamma_{mN}$ it follows that had~$\Gamma_{e\text{N}}$ not surrounded the north cycle then
\be\label{GammaeNZEROAREA}
\int_{\Gamma_{eN}} (A_a)_\mu dx^\mu\,\to\,\,0\
\ee
in its zero-area limit. Repeating the same analysis but for the south cycle $\Gamma_{e\text{S}}$ one reaches
\be\label{Electric2}
\int_{\Sigma_{\Gamma_{e\text{S}}}}\, (F_a)_{\theta t_E}\,\diff\theta\,\wedge\, \diff t_E  \,\limSouth \,\Bigl(v_{1a}\,-\,\frac{n}{2m}\, v_{2a}\Bigr)\,,
\ee
and again, had~$\Gamma_{e\text{S}}$ not surrounded the south pole then
\be\label{GammaeSZEROAREA}
\int_{\Gamma_{eS}} (A_a)_\mu dx^\mu\,\to\,\,0\,.
\ee
in its zero-area limit. 
When~$m$ and~$n$ correspond to orientable~$\Sigma_{e\text{N}}$ and~$\Sigma_{e\text{S}}$, then starting from relations~\eqref{orientable} and applying Stokes theorem one obtains analogous equations to~\eqref{Electric1}~-~\eqref{GammaeSZEROAREA}.~\footnote{Note that the form of the fixed-points~\eqref{SolsCardy} implies that the integral along the twisted time cycle vanishes at~$\tau\,\to\,-\frac{n}{m}$
\be
\int_{\Gamma} \,{A_a}_\mu\, \diff x^\mu\,=\, v_{1a}+\tau v_{2a}\,\liMN\,0\,.
\ee
}
From~\eqref{Electric1},~\eqref{GammaeNZEROAREA},~\eqref{Electric2}, and~\eqref{GammaeSZEROAREA} it follows that
\be\label{ElectricField}
(F_a)_{\theta t_E}\,=\, \frac{ 1 }{2\beta}\,\frac{n\, (a-\widehat{a})}{N}\, \Bigl(\delta(N)\,-\,\delta(S)\Bigr)\,-\,\Ethi\,.
\ee
~$\Ethi$ stands for integrable contributions that we have given a name for latter convenience (See footnote~\footnote{In appendix~\ref{app:Charge} we will use these terms to clarify why there is no condensation of electric charge at the fixed loci of the rotational action even if there is a non-vanishing distributions for the charge density located at such position.}). As the supersymmetric path integral in the punctured space~$\mathcal{I}^{\times\times}$, localizes to flat connections, then~$\Ethi\=0\,$.

\paragraph{Similarities with Gukov-Witten surface operators}{The boundary conditions~\eqref{FthhetaPHi}, \eqref{ConditionMagSouth}, \eqref{Electric1}, \eqref{Electric2} are reminiscent of the ones used in~\cite{Gukov:2006jk} to define surface operators. In particular, of the choice~$\alpha\neq 0$,~$\beta=\gamma=0$ in~(2.2)~\cite{Gukov:2006jk}.

The constituents of the~$(m,n)$ bound-states (as defined by the path integral~$\mathcal{I}^{\times\times}$ in the presence of punctures) are reminiscent of the surface operators defined by the~$(\alpha, \beta=0,\gamma=0)$-boundary conditions of~\cite{Gukov:2006jk}. In the case here studied, the connection~$A$ has more than one non-vanishing components fixed. It would be interesting to explore if there is a relation in between the~$S$-dual pair~$(\alpha,\eta)$ discussed in~\cite{Gukov:2006jk} and the eigenvalue components~$(v_1,v_2)$ here discussed.

The Dyonic fluxes~(\eqref{FthhetaPHi}+\eqref{ConditionMagSouth},\eqref{ElectricField}) (say at~$N$) span a lattice descending from the one in which the couple~$(v_{a1},v_{a2})$ lives in. This should be implemented generalizing ideas of~\cite{Gukov:2006jk} (subsection~\emph{Extensions of Bundles}). For the present purposes, we do not need to go into such construction. Instead, we have assumed that such an extension of bundle exists; and chosen to focus on a specific patch, on which the Dyonic fluxes take the forms~(\eqref{FthhetaPHi}+\eqref{ConditionMagSouth},\eqref{ElectricField}).

\paragraph{Infinite Noether charges} There is a difference between the surface condensates  here discussed, and what is called~\emph{a monodromy defect}: The former emerge when the geometric background (including background gauge fields) on which the UV theory is placed upon is deformed to specific set ups ($\tau\to -\frac{n}{m}$). The latter, instead, are defined by an explicit deformation of the fundamental path integral.

Note that the Noether charges of these bound-states are infinity.~\footnote{They involve integrals of a product of coincident Dirac deltas with constant test functions.} Regularizing these infinities would require modifying the UV theory.~\footnote{One would need to insert the~\emph{inverse} monodromy defect in the fundamental theory. Meaning by that, a (supersymmetric)  defect that would cancel the monodromy charges of the bound-state, however, such an insertion would also be perceived by every other observable in the theory, and thus it would be a meaningful modification of the UV theory, and we do not wish to do so.} Thus, we are draw to accept that the infinite Noether charges of the~$(m,n)$ bound-states are not a misunderstanding, but a feature of these configurations.~\footnote{The fact that these operators have infinite Noether charges, and are charged under~$\mathbb{Z}^{(1)}_N$-form symmetry, makes them potential candidates for~\emph{redundant operators/states} in~$4d$~$\mathcal{N}=4$ SYM. By redundant we refer to the definition given in~\cite{SeibergSeminarSCGP}: although they have infinite energy in the fundamental UV theory they remain as fundamental degrees of freedom in the effective field theory of the ordered phases as argued in section~\ref{TopOrder}.}

\subsection{Electromagnetic~$\mathbb{Z}^{(1)}_N$ one-form charge  of phases: An order parameter}
\label{OneFormSymmetry}
The goal of this section is to explain how the~$(m,n)$ phases are characterized by a doublet of $\mathbb{Z}^{(1)}_{N}$ one-form charges: the \emph{electromagnetic~$\mathbb{Z}^{(1)}_{N}$ one-form charge of the phase}. This charge is an order parameter in the sense that it specifies a single phase among the family of phases coming from the fixed points that we have focused on~\eqref{SolutionsMNPhases}.

The set of solutions~\eqref{SolutionsMNPhases} includes fixed-points that are equivalent under Weyl transformations and contains more solutions than the ones dominating as~$\tau\to-\frac{n}{m}\,$, i.e., the ones that characterize the~$(m,n)$-phase. The gauge orbits intersected by the potentials associated to~\eqref{SolutionsMNPhases} : 
\be\label{CartanAnsatzFinal}
A^{(\widehat{p},\,\widehat{q})}_{(c,\,d)}\=\sum_{a\=1}^{N\,-\,1}\,\Bigl(\frac{a\,d\,+\,\widehat{p}}{N} \,\,\diff t_E \,-\,\frac{a\,c\,+\,\widehat{q}}{N} \,\diff \phi\Bigr)\, T_a\,, \quad \phi\=\frac{1}{2}(\phi_1\,+\,\phi_2)\,,
\ee
\footnote{The conventions used in this subsection are summarized in appendix~\ref{Nality}.} form a representation of a~$\mathbb{Z}^{(1)}_N\times \mathbb{Z}^{(1)}_N$ one-form transformation at any fixed~$(\widehat{p},\widehat{q})$. For reasons that will be clear below, these transformation will be called electromagnetic~$\mathbb{Z}^{(1)}_N$ one-form symmetry.

The Cardy-like expansion~$\tau\to-\frac{n}{m}\,$ selects only solutions with~$(c,d)=(m,n)$ within~\eqref{CartanAnsatzFinal}. These are
\be\label{SolsCardy2}
v_a\= \frac{(a-\widehat{a})\,\text{mod}\,N}{N}\, (m \tau\,+\,n)\,.
\ee
~$\widehat{a}\,=\,0,\ldots N-1\,$.  These~$N$ leading fixed points carry the very same electromagnetic~$\mathbb{Z}_N^{(1)}$ one-form charges. We recall that in this note, we are assuming~$N$ to be prime, but our conclusions can be generalized to other values of~$N$. 

The Lie algbra generators~$T_a$ in~\eqref{SolutionsMNPhases} are assumed to be in any of the~$N$ fundamental representations of the Lie algebra of~$SU(N)$: the defining representation together with its fully anti-symmetrized tensor product representations.

\emph{Electric~$\mathbb{Z}^{(1)}_N$ one-form charge:} Let us explain what we mean by electric~$\mathbb{Z}^{(1)}_{N}$ one-form charge. Let us start with the subsector~$\widehat{p}\=\widehat{q}\=0\,$. That this subsector carries a representation of the center (or electric)~$\mathbb{Z}_N$ symmetry follows from the fact that for~$\widehat{p}\=0$ the time component of~\eqref{CartanAnsatzFinal} is generated by the maps
\be\label{SingularElectric}
G_{el}(t_E;d) \equiv e^{2\pi\i\,d\,\frac{t_E}{\beta}\,\sum_{a\=1}^{N\,-\,1}\, \frac{a}{N}\, T_a} \,, \qquad n\=1\,,\,\ldots\,N \,\text{mod} \,N\,.
\ee
When one moves along the Polyakov cycle once,~$G_{el}$ interpolates between the unit element at~$t_E=0$ and an element of the center of~$SU(N)$
\be\label{ElectricZN}
G_{el}(\beta;d)\= e^{2\pi \i \frac{d}{N} n_{\textbf{r}}}\,\, \mathbb{I}\,.
\ee
where~$n_{\textbf{r}}$ is the~$N$-ality of the fundamental representation~$\textbf{r}$.

Thus, in particular, the~$(m,n)$ bound-state (with~$\widehat{p}\=\widehat{q}=0$) carries discrete charge~$n$ under the center (electric one-form)~$\mathbb{Z}^{(1)}_N$ symmetry. Using an obvious generalization of~\eqref{SingularElectric}, one can interpolate from the root of identity~\eqref{ElectricZN}, which is the one associated to the phase~$(m,n)$ (with~$\widehat{p}\=\widehat{q}=0$), to the root of the identity associated to any other phase~$(c,d)$ (with~$\widehat{p}\=\widehat{q}\=0$).

\emph{Magnetic~$\mathbb{Z}^{(1)}_{N}$ one-form charge:} Equivalently, the space-like component of~\eqref{CartanAnsatzFinal} is generated by 
\be
G_{mag}(\phi;c) \equiv e^{-\i \,c\,\phi\,\sum_{a\=1}^{N\,-\,1}\, \frac{a}{N}\, T_a} \,, \qquad c\,\text{mod} \,N\=1,\ldots\,N\,,
\ee
with~$\phi\equiv\frac{\phi_1+\phi_2}{2}\,$. Analogously, when one moves along the magnetic $\phi$-cycle once, the map~$G_{mag}$ interpolates between the unit element at $\phi=0$ and an element in the center of~$SU(N)$
\be
 G_{mag}(2\pi;m)\= e^{-2\pi \i \frac{c}{N} n_{\textbf{r}}}\, \mathbb{I}\,.
\ee
Thus the conclusions are analogous to the ones of the electric case and we do not repeat them.

\emph{Generic values of $\widehat{p}$ and~$\widehat{q}$ :} For~$\widehat{p}\,, \,\widehat{q}\,\neq\,0$ the generalization of the above conclusion is simple. In that case the time and angular components of the gauge potential~$A$ associated to a given~$(c,d)$ fixed-point, do not exponentiate to a root of unity, but the difference between the components associated to a gauge potential~$A$ of any two such fixed-points exponentiate to a root of unity. This means that one can move between fixed-points, at fixed~$\widehat{p}$ and~$\widehat{q}$, by using the obvious generalizations of the maps~$G_{el}$ and~$G_{mag}$ i.e. those that implement the changes in discrete charges~$\Delta c$ and~$\Delta d$ from one (family of~$N$) fixed-point to another, as one winds around the relevant cycle once.

At last, as discussed around~\eqref{Identification0}, some solutions in~\eqref{CartanAnsatzFinal} are identified under Weyl permutations. This is the same as saying that part of the electromagnetic~$\mathbb{Z}^{(1)}_{N}\,\times\,\mathbb{Z}^{(1)}_{N}$ discrete symmetry group is gauge. The physically relevant global discrete charges being
\be
(c,d)\,\in\,\mathbb{Z}^{(1)}_{N}\,\times\,\mathbb{Z}^{(1)}_{N}\,/\,\mathcal{L},
\ee
where~$\mathcal{L}$ was defined in~\eqref{Identification0}. In short, the division by~$\mathcal{L}$ is equivalent to disregarding non co-primes~$c$ and~$d\,$. The co-prime~$c$ and~$d$ are called magnetic and electric $\mathbb{Z}^{(1)}_N$ one-form charges of the phases, respectively.

\section{Brief summary and questions for the future}
\label{sec:5}

We have shown that in the zero-temperature limit where the thermal partition function of~$SU(N)$~$\mathcal{N}=4$ SYM (at any coupling) reduces to the superconformal index, topological order emerges when a certain complex parameter~$q=e^{2\pi\text{i}\tau}$ related to the rotation of the spatial part of the geometry (where the theory is quantized upon) is taken to a root of unity. 
Each of these phases is characterized by~$N$ {bound-states} made out of two surface components. The two component operators are localized at the fixed loci of the rotational action. The fixed loci are two two-tori located at north and south of the~$S_3$, respectively. Both components carry localized electromagnetic flux, this is, when the flux is measured by two independent families of probe fundamental Wilson-line operators, the measured holonomy remains independent of the area enclosed by the lines. 

These bound states are classified by their charge under a~$\mathbb{Z}^{(1)}_{N}\times \mathbb{Z}^{(1)}_{N}$ one-form symmetries~\cite{Gaiotto:2014kfa}. The two discrete~$\mathbb{Z}^{(1)}_{N}$ one-form symmetries in the tensor product are called magnetic and electric, respectively, because they act upon the fundamental Wilson loops of~$SU(N)$ when the latter rest along time-like and space-like cycles, respectively.~\footnote{Only part of this symmetry is global i.e.~$\mathbb{Z}^{(1)}_{N}\times \mathbb{Z}^{(1)}_N/\mathcal{L}\,\in\,\mathbb{Z}^{(1)}_{N}\times \mathbb{Z}^{(1)}_{N}$ where the operation~$\mathcal{L}$ was defined in~\eqref{Identification0}.}  Like the magnetic (resp. electric) flux, the magnetic (resp. electric) discrete charge of the~$(m,n)$ bound-state is determined by the integer~$m$ (resp.~$n$). The magnetic (resp. electric) discrete one-form symmetry acts over the flux charge~$m$ (resp.~$n$), transforming bound-states into bound-states. The Cardy-like limit projects the representation of the electromagnetic one-form symmetry to a subset of~$N$ solutions with well-defined set of electromagnetic one-form charges specified by two co-prime integers~$m$ and~$n$. The electromagnetic one-form charge is the order parameter of the specific~$(m,n)$ quantum phase. 

The bound states that superpose to form these phases are a one-dimensional uplift of Anyons. In particular, based on classical results in the literature we have argued that the operators creating the physical excitations in the effective theory at the a given~$(m,n)$ phase form a Verlinde fusion algebra (section~\ref{sec:QuantumGroup}).

There are some questions and problems that we have decided to leave for the future to study. Some of these are:

\begin{itemize}

\item To find the explicit relation between the approaches described in the subsection~\ref{sec:QuantumGroup} and subsection~\ref{sec:FiberingOperator}. In particular, it would be interesting to understand how the Verlinde algebra among the operators $\mathcal{O}_{\textbf{r}}$ arises from the Cardy-like expansion of the description in terms of the fibering operator~$\mathcal{F}$ on~$T^{\tau}_2\times S_2$. Notice that doing so, requires the extraction of the leading Casimir pre-factor and a careful analysis of subleading corrections in the limit~$\tau\to-\frac{n}{m}$, along the lines of the one presented in the companion paper~\cite{PaperBulk}.~\footnote{Similar results, at least in spirit, have been already reported in the literature~\cite{Benini:2015noa} about 3d A-twisted gauge theories.} We will address this problem in forthcoming work. 

\item Does a larger quantum group structure exist at finite values of~$\tau$? Such a possibility is suggested by the naive observation that one can define a quantum group~$U_q({SU}(N))$ for every loop~$\ell$ in~$T_2$. It would be interesting to identify such larger structures, e.g., it is plaussible that one could find structures reminiscent of the ones studied in~\cite{Costello:2013zra,Costello:2018gyb,Costello:2017dso,Witten:2016spx}.

\item What is the interpretation of the~$\mathbb{Z}^{(1)}_N$ symmetry in the AdS side of the duality? These symmetry transformations should leave invariant certain subsets of the (Euclidean) gravitational solutions studied in~\cite{Aharony:2021zkr}. It would be interesting to study this problem in more details.

    \item Can the interpretation of~$(m,n)$ phases as a mixed state of~$N$ bound states~\footnote{To these phases one can associate the density matrix
\be\label{DensityMatrix}
\rho^{G/G}\,=\,  \sum_{\widehat{a}=1}^N |\widehat{a} \rangle \langle \widehat{a}|
\ee
with~$|\widehat{a}\rangle$ ($\widehat{a}=1,2,\ldots, N$), being a ket associated to the~$\widehat{a}$-th bound-state.~{We assume the normalization~$\langle \widehat{a}|\widehat{a} \rangle=1\,$}, then
\be
\mathcal{I}^{(1)}\,\leftrightarrow\, \text{Tr} \rho^{G/G} \,=\, \sum_{\widehat{a}=1}^N  \langle \widehat{a}|\widehat{a} \rangle =N\,.
\ee} be helpful to understand universal near zero-temperature corrections of the thermal partition function in expansions other than~\eqref{BPSLimit1}?~\footnote{As the superconformal index is protected under variations of the gauge coupling our conclusions should remain valid at any value of the gauge coupling. It would be interesting to check this robustness coming from observables that depend on the gauge coupling. In order to do this, integrability tools can be useful.}

\item Can the imaginary part of~$\tau$ (considered to be non-vanishing due to convergence reasons) be re-interpreted as the~$\eta$ discussed in~\cite{Gukov:2006jk}. If yes, then the conclusions in this paper can be translated to a set-up with a real background geometry defined as~$\tau=-\frac{n}{m}\,$.~{If that is the case then it would be interesting to aim for a deeper understanding of the relation with the framework studied in~\cite{Kapustin:2006pk,Gukov:2006jk}\cite{Frenkel:1995zp} and other relevant references.}

\item Generic four dimensional~$\mathcal{N}=1$ superconformal indices are known to have a factorized expression in terms of vortex and anti-vortex blocks, each of them localized at the north and south loci of the rotational action~\cite{Yoshida:2014qwa,Peelaers:2014ima,Nieri:2015yia}. A similar type of formula has been put forward in~\cite{Goldstein:2020yvj}. It would be interesting to find the relation between such a formula and the Bethe ansatz one~\eqref{BARepIndex}.~\footnote{Understanding such a relation (which does not seem to be trivial) would probably hint at the form of the effective action of~$\mathcal{B}$ at finite~$\tau\,$.}

\item One could ask if non-supersymmetric gauge theories could exhibit zero-temperature topologically ordered phases similar to the ones here argued. We note that phase structures with different physical properties, but with seemingly related mathematical roots have been identified in varied contexts~\cite{tHooft:1981bkw,Cardy:1981qy,Cardy:1981fd,Shapere:1988zv,Seiberg:1994rs,Intriligator:1995id}.

\end{itemize}

\section*{Acknowledgements}
~It is a pleasure to thank D. Anninos, A.A.Ardehali, P. Benetti Genolini, S. Murthy, and S. Schafer-Nameki for useful conversations and/or discussions. This work was supported by the ERC Consolidator Grant N. 681908, “Quantum black holes: A microscopic window into the microstructure of gravity”.

\appendix

\section{Universal affine Kac-Moody algebras from four-dimensional gauge transformations}
\label{sec:4}
In this section, we show how the gauge transformations of a 4d gauge theory~\footnote{... for which physical observables localize into integrals over the phase space of 4d locally flat connections...} induce an affine Kac-Moody algebra per one-cycle~$\ell$ for the effective theory at the punctured boundary.

\subsection{Chiral current algebras from gauge transformations}\label{SUNKM}

The tangential component of the fundamental gauge potential along the cycle~$\ell$ at the puncture~$\times$ is
\be
{A}_{\ell} \,\equiv\,  \frac{\diff x^\mu_{\ell}}{\diff \a}\, {A}_{\mu}\,\bigl(x^\mu_{\ell}\bigr)\,=\,A_{\gamma\,=\,\ell}\,.
\ee
This can be understood as a linear combination of the two components in the boundary conditions~$A_\gamma$ defined and used in the introductory section~\ref{sec:MainIdea}.
Take now a smooth map~$G$ from the geometry~\eqref{background_metric0} onto~$SU(N)$, and denote its restriction to the cycle~$\ell$ in~$\times$ as~$G \Bigl|_{\ell}=G_\ell(\alpha)$. The gauge transformation induced by the group element~$G$ upon~${A}_\mu$, when restricted to the cycle~$\gamma$, is
\be\label{Transformation}
\delta_{G_\ell} {A}_{\ell}\= -\i G_\ell^{-1}\partial_{\alpha} G_\ell\,+\, G_\ell^{-1} {A}_{\ell} G_\ell\,-\,{A}_{\ell}\,.
\ee
If we focus on configurations~$A_{\gamma=\ell}$ that are locally pure gauge, i.e., pure gauge in the punctured spacetime (these include the connections associated to the complex fixed-point configurations~\eqref{CartanAnsatzFinal}), then one can decompose~$A_\ell$ in components
\be\label{currentDef}
A^a_{\ell}(\alpha)\,=\,\frac{1}{N}\,\text{Tr}\Bigl( X^a\,{A}_{\ell}(\alpha) \Bigr)\,.
\ee
where~$X^a$ is an element of the Lie algebra of~$SU(N)\,$ and Tr is the trace in the adjoint representation.

From the transformation~\eqref{Transformation} and the decomposition~\eqref{currentDef} it follows that
\be\label{variationJ}
\delta_{G_\ell} A^a_\ell \,=\,-\frac{\i}{N}\, \,\text{Tr}\Bigl(X^a\,G_\ell^{-1}\partial_{\alpha} G_\ell\Bigr) \,+\, \frac{1}{N}\,\text{Tr}\,\Bigl(\bigl(G_\ell X^a\,G_\ell^{-1}-X^a\bigr)\,\widetilde{A}_{\ell}\Bigr)\,.
\ee
Acting with a gauge transformation induced by a generic group element~
\be
G_\ell\=\exp(-\i \sum_{b}\varepsilon_b X^b )\,,
\ee
with infinitesimal Lie algebra parameter~$\varepsilon\,$, on~$J^a_\ell\,$, then~\eqref{variationJ} takes the form
\be\label{VariationKM}
\begin{split}
\delta_{\varepsilon} A^a_\ell(\alpha) & \= -\frac{1}{N}\, \,\partial_{\a}\varepsilon_b \,\text{Tr}\Bigl(X^a\,X^b\Bigr) \,-\frac{\i}{N}\,\varepsilon_b\, \text{Tr}\Bigl( [X^a\,,X^b]\,A_{\ell}(\alpha)\Bigr)\, \\
&\=-\,\frac{1}{N}\, \partial_{\a}\varepsilon_b \,\text{Tr}\Bigl(X^a\,X^b\Bigr)\,+\,\varepsilon_b f^{a b}_{\,\,\,\,c}\, J^c_\ell(\alpha)\,,
\end{split}
\ee
which means that a variation of~$J_{\ell}^a$ induced by motion along a gauge orbit is generated by a Kac-Moody algebra action at level~$k=1\,$.  In a phase space \underline{of locally flat connections} the variation~\eqref{VariationKM} can be thought of as induced by the Poisson bracket action of~$J_\ell$ upon itself
\be\label{KMalgebra}
-\i [A^a_{\ell}(\alpha), A^b_{\ell}(\alpha^\prime)]\= -\, \delta^{a b}\,\delta_{per}^{\prime}(\alpha-\alpha^\prime) \,+\,\delta_{per}(\alpha-
\alpha^\prime)\,f^{a b}_{\,\,\,\, c} A^c_\ell(\alpha)\,.
\ee
The symbol~$\delta_{per}$ denotes the periodic delta function with period~$2\pi\,$. The factor of~$N$ that cancels the~$\frac{1}{N}$ in the central extension, comes from the fact that~$X^a$ is in the adjoint of~$SU(N)$. In that case~$(X^a)_{bc}\=-\i f_{a b c}$, the identity~$f_{a c d}f_{b c d}\= N \delta^{a b}$ implies
\be
\text{Tr}\Bigl(X^a\,X^b\Bigr)\= N \,\delta^{a b} \,.
\ee
The same analysis can be repeated for any other cycle~$\ell$: note that the level does not depend on the selection of~$\ell$.

\subsection{On the binding between North and South edge modes}\label{sec:WZNW}

As explained in the main part of the paper
\be\label{IndexFundamentalExp2}
\mathcal{I}\underset{\tau\,\to\, 0}{\simeq} e^{\text{Casimir-type contribution}} Z_{G/G} \,,
\ee
where~$Z_{G/G} \,=\, \sum_{\widehat{a}=1}^{N} 1\,=\,N$ is the partition function of the gauged~$SU(N)_1$ WZNW model on $T_2\,$~\cite{Gawedzki:1988nj,Karabali:1989dk,Witten:1991mm,Spiegelglas:1992jg}~\cite{Witten:1991mk} .

The~$G/G$ WZNW model can be quantized via BRST-like~\cite{Karabali:1989dk,Spiegelglas:1992jg} or path integrals~\cite{Gawedzki:1988nj,Witten:1991mm}~\cite{Witten:1991mk} methods. The path integral representation of the partition function of such model can be factorized~\cite{Witten:1991mm} and written in the form
\be\label{Factorization}
\begin{split}
Z_{G/G}&\,:=\,\int \text{D}A \text{D}B\, \overline{\chi(A,B)}\,{\chi(A,B)} \\
&\= \,\# \,\text{of conformal primaries at genus one}\,=\, N
\end{split}
\ee\footnote{The second line in this equation can also be shown using the relation between~$G/G$ WZNW on a two-torus and~$G$ Chern-Simons theory on a three-torus~\cite{Blau:1993hj}~\cite{Verlinde:1988sn,Witten:1988hf,Moore:1988qv}. }
where~$\chi$,~$\overline{\chi}$ are the holomorphic and anti-holomorphic wave functions defined in~\cite{Witten:1991mm} for~$G=SU(N)_1$
\be\label{WaveFunctions}
\chi\,:=\,\int \text{D}G_N\, e^{- I(G_N;A,B)}\,, \qquad \overline{\chi}\,:=\,\int \text{D}G_S\, e^{- I^\prime(G_S;A,B)}\,.
\ee
\footnote{
\be
\begin{split}
I(G;A,B)&\,:=\, I(G)\,+\,\frac{1}{2\pi} \int \text{d}^2 \alpha \,\text{Tr} A_2 G^{-1}\partial_{\alpha_1} G \,-\,\frac{1}{2\pi} \int \text{d}^2 \,\alpha \text{Tr} B_1 \partial_{\alpha_2} G G^{-1} \\ 
&\qquad \qquad +\frac{1}{2\pi}\int \text{d}^2 \alpha \,\text{Tr} B_1 G A_2 G^{-1}\,-\,\frac{1}{4 \pi}\,\int\,\text{d}^2\alpha\, \text{Tr}(A_1 A_2 +B1 B_2)\,, \\
&\Bigl(\text{and} \qquad I^\prime(G;A,B)\,:=\,I(G;B,A) \Bigr)\,,
\end{split}
\ee
is the action of the WZNW model on a two-torus with world line coordinates~$\alpha_1\sim\alpha_1+2\pi$ and~$\alpha_2\sim \alpha_2+2\pi$ and line element~$ \text{d}\alpha_1 \text{d}\alpha_2\,$, in the presence of background gauge fields~$A:=A_1 \text{d}\alpha_1+A_2 \text{d}\alpha_2$ and~$B:=B_1 \text{d}\alpha_1+B_2 \text{d}\alpha_2$ dual to the commuting chiral currents whose origin will be motivated below.~The action of the undeformed WZNW model being
\be
I(g)\,:=\, -\,\frac{1}{8 \pi} \int \text{d}^2 \alpha \text{Tr}\Bigl(G^{-1} \partial_{\alpha_1} G \cdot  G^{-1} \partial_{\alpha_2} G\Bigr) \,-\,\i \Gamma(G)
\ee
with the Wess-Zumino term~$\Gamma(G)$ defined in a three-dimensional manifold with a two-torus (spanned by~$\alpha_1$ and~$\alpha_2$) as boundary~\cite{Witten:1991mm}.

}
~$A=A_\ell$ and~$B=B_\ell$ are gauge fields for the two chiral symmetries that were previously identified with the symmetries generated by the holomorphic and anti-holomorphic currents~$J_{\ell}$ at each of the punctures. Then the two wave functions in~\eqref{Factorization} can be thought of as path integrals over the north and south edge modes degrees of freedom (in the presence of gauging). At last, an expression for~$W_2$ can then be obtained by plugging~\eqref{WaveFunctions} in~\eqref{Factorization} and the result in~\eqref{IndexFundamentalExp2}, and comparing the answer with~\eqref{RelationFunctionalTwoP}.

\subsection{Polyakov loop correlators in Cardy-like expansion}
The localization procedure reduces gauge potentials to Cartan holonomy variables
\be
A_\mu, A_\gamma \longrightarrow  u_i \sim u_i\,+\,1\,.
\ee
Moreover, for a supersymmetric choice of cycle~$\ell$ (for instance the time-cycle that we will define below as~$\Gamma$) the Polyakov loop operators in representation~$\textbf{r}$ localize to characters of the gauge group
\be
\text{Tr}M_{\ell| } \,\longrightarrow\, \chi(u)
\ee
with~$\chi(u):=\sum_{\rho \in \textbf{r}} \e(\rho(u)) $ being the character of the representation~$\textbf{r}$~of~the gauge group.

 The path integral representation localizes to the following expression, 
 \be\label{TheIndex}
\mathcal{I} = \kappa\, \int_0^1 \prod_{i=1}^{N-1} du_i   \prod_{i > j =1}^{N}\bigl( \theta_0(u_{ij})\,\theta_0(u_{ji})\bigr)\,\prod_{i\neq j=1}^{N}\prod_{a=1}^3 \Gamma_{\text{e}}(u_{ij} + \Delta_a;\tau)\,,
 \ee
 which matches the Hamiltonian representation of the index. The constant~$\kappa$ and the elliptic functions in the argument are the ones given in the introduction of the companion paper~\cite{PaperBulk}.
 
 The expectation values of Polyakov loops~$\prod_{\textbf{r}} M_\ell$ localize to
\be\label{MCorrelatorsPolyakov}
\frac{1}{\mathcal{I}}\,\kappa \,\int_0^1 \prod_{i=1}^{N-1} du_i  \prod_{i > j =1}^{N}\bigl( \theta_0(u_{ij})\,\theta_0(u_{ji})\bigr)\,  \prod_{i\neq j=1}^{N}\prod_{a=1}^3 \Gamma_{\text{e}}(u_{ij} + \Delta_a;\tau)\, \prod_{\textbf{r}} \chi(u)\,.
 \ee
The previous arguments imply that in Cardy-like expansion (around~$\tau\to0$) the correlators~\eqref{MCorrelatorsPolyakov} must be re-writable as correlators of the gauged $SU(N)_1$ WZNW theory on~$T_2\,$, which have the form~\cite{Blau:1993tv}
\be\label{CorrelatorsSUN}
\begin{split}
\sum_{j\,=\,0}^{N-1}\,\mathcal{X}(\chi(\phi^{(j)}_i)) \prod_{a} \chi_{\textbf{r}_a}(\phi^{(j)}_i)\,.
\end{split}
\ee
In these expression~$\mathcal{X}$ is some function of characters of~$SU(N)$. The~$\phi^{(0)}_{i}=0$ and the configurations over which the correposponding integral localizes over (the~$L$'s before) are~\cite{Blau:1993tv}
\be
\phi^{(j)}_i(n)\,:=\, \frac{1+\delta_{i,j}}{k+N}\,,\qquad i,j\,=\,1,\ldots, N-1\,.
\ee
\footnote{It seems possible to check this prediction from scratch, we will try to do this elsewhere.} Similarly to~$SU(N)$, for the~$U(N)$ index we expect the set of Polyakov loop correlators to be equivalent to correlators of the~$U(N)_k/U(N)_k$ WZNW model. The~$U(N)$ ones are obtained from the~$SU(N)$~\eqref{CorrelatorsSUN} by exchanging the characters of~$SU(N)$ by the~$U(N)$ ones. In this case the configurations~$\phi^{(j)}_{i}$ over which the sum runs over are determined by the folllowing equations, (which we borrow from~\cite{Okuda:2012nx})
\be
(N\,+\,k)\, \phi^{(j)}_a -\sum_{b=1}^{N} \phi^{(j)}_b+\frac{N-1}{2} - j_a\,=\, 0\,, \qquad j_a \,\in\,\mathbb{Z}\,,
\ee
with the equivalence relation~ $\phi^{(j)}_i\,\sim\, \phi^{(j)}_i\,+\,1$ in mind. Assuming~$N$ prime,~ and using level-rank duality these equations map to~\cite{Okuda:2012nx}\cite{Naculich:2007nc}
\be\label{RootsOfUnity}
N\, \phi^{(j)} - j\,=\, 0\,, \qquad j \,\in\,\mathbb{Z}\,.
\ee
which naturally correspond to~$N$-th roots of unity.
 These configurations solve the Bethe ansatz equations of the gauged~$U(1)_N$ WZNW model which is level-rank dual to the gauged~$U(N)_1\,$ model, and thus are equivalent in the infrared~\cite{Closset:2017bse} i.e. they have the same quantum group symmetry. The roots~\eqref{RootsOfUnity} are predicted to be in one-to-one relation with the gauge orbits~$L$'s (Bethe roots) that dominate the Cardy-like limit of the~$U(N)\,$ index. They need not to look the same: theories that are equivalent (duals) can have Bethe roots with different forms.~We will comeback to identify thee roots in section~\ref{sec:WZNW} (See also the last paragraph in appendix~\ref{app:Integrability}).

\section{Center elements from the defining representation of~$SU(N)$} \label{Nality}

Let us ask for the set of~$N-1$ real~$v_{a}$'s solving
\be\label{CenterEquations}
\exp\bigl( 2\pi \i \,\sum_a v_{a}\,g^{-1} T^a g\bigr) \= X_{(\ell)} \, \mathbb{I}_{N\times N}\,, \qquad X_{(\ell)}\=e^{2\pi \i \frac{\ell}{N}},
\ee
with~$\ell\,\in\,\mathbb{Z}\,$. In this equation~$g$ is an arbitrary element of~$SU(N)$, and~$T_a$ is the defining representation of Lie algebra of~$SU(N)$. The other fundamental representations are the fully antisymmetrized tensor product of the defining one. As said before, in this paper we  assume~$N$ to be prime. Note that the solutions for the numbers~$X_{(\ell)}$ is independent of~$g$, however, the form of the components~$v_{a(\ell)}$ does depend on the choice of~$g\,$.

If we fix the basis for the defining representation as
\be\label{Cartan}
T_{k}\= e_{k\,,\,k}\,-\,e_{k\,+1,\,k+1}\,,\qquad k\=1\,,\ldots\,, N-1\,,
\ee
with~$e_{i\,,\,j}$ denoting the~$N\times N$ matrix with unique non-vanishing entry~$1$ in the~$(i,j)$-th position, the equations~\eqref{CenterEquations}, for a fixed~$\ell=1,\ldots N$, are solved by
\be\label{solutionsCenteruK}
\begin{split}
e^{2 \pi\i v_{a(\ell)}} &\= \bigl(X_{(\ell)}\bigr)^a\,, \\
v_{a(\ell)} &\= \ell \,\frac{a}{N}\, \text{mod}\, \mathbb{Z} \,,\qquad  \ell\=1\,,\ldots N\,.
\end{split}
\ee
We note that in the basis~\eqref{Cartan} the Cartan matrix takes its canonical form i.e.~$2$ in the diagonal elements and~$-1$ in the main off-diagonal elements. As said before, should we have chosen a different basis for the~$T_a$'s, related to~\eqref{Cartan} by a similarity transformation, the~$X_{(\ell)}$ remains the same but the form of the $v_{a(\ell)}$'s changes. For the~$N$ fundamental representations~$\textbf{r}$ we will always use the  basis obtained from antisymmetrized tensor products~\eqref{Cartan}.

\section{Chromoelectric charge?}\label{app:Charge}

Let us show how there is no accumulation of electric charge in the two components of the~$(m,n)$ bound-states, even though the charge density is a non-vanishing distribution at north and south loci of the rotational action. There are only two choices of~$\Ethi$ in~\eqref{ElectricField}
\be\label{Etheta}
-\Ethi\,\equiv\, 0 \qquad \text{\bf{or}}  \qquad \,\propto\,\frac{q^{a}_{bulk}}{\sqrt{\text{detg}(\theta)}}\,,\,\qquad \sqrt{\text{detg}(\theta)}=\frac{r_{S_3}^3}{2}\sin(2\theta)\,, 
\ee
~\footnote{ In these expressions we have reinstated the radius of~$S_3$.} for which the distribution of chromoelectric charge can be localized at north and south i.e. for which
\be\label{GaussLaw}
\begin{split}
\rho^a_e&\,\equiv\,  \nabla^\mu \Bigl( F^a_{\mu t_E}\Bigr)\\&= 2\,\Bigl(\cot (2\theta)\Bigr)\,\frac{1}{r_{S_3}^2}\, F^a_{\theta t_E}\,+\,\frac{1}{r_{S_3}^2}\,\partial_\theta F^a_{\theta t_E} \\&\= 2\,\Bigl(\cot (2\theta)\Bigr)\, \frac{(q^a+q^a_{bulk}) \delta{(N)}\,-\,(q^a+q^a_{bulk}) \delta{(S)} }{(2\pi)^2}\,+\, \frac{q^a \delta^\prime{(N)}\,-\,q^a \delta^{\prime}{(S)} }{(2\pi)^2} \,,
\end{split}
\ee
with~$q^a \propto \frac{n \,a}{N}\,\frac{r_{S_3}}{2\beta}\,$, is made of distributions localized at the north and south loci.
The second choice in~\eqref{Etheta}, which implies condensation of electric charge, is both non-integrable and not locally flat, thus we can take it to be zero. Such contribution is the electric field generated by an homogeneously charged string carrying total electric charge~$q^a_{bulk}$. We have kept this contribution to make clear how~$q^{a}$ can not be interpreted as electric charge in equation~\eqref{GaussLaw}.

Indeed, assuming~$q^a_{bulk}\,=\,0$, and using Gauss law~\eqref{GaussLaw} to compute the electric charge enclosed by a portion of the three sphere $t_E=\text{fixed}$ within~$0<\theta<\theta_0$, one obtains
\be\label{LocalizationCharges}
\begin{split}
\int^{\theta_0}_0 d\theta \int_0^{2\pi} d \phi_1 \int_0^{2\pi} d\phi_2 \sqrt{\text{detg}}\, \rho^i_e &\= \int^{\theta_0}_0 d\theta \int_0^{2\pi} d \phi_1 \int_0^{2\pi} d\phi_2\,\sqrt{\text{detg}}\, \nabla^\mu \Bigl( F^i_{\mu t_E}\Bigr)\\&
\= 0\,.
 \end{split}
\ee
Thus, although the density of charge is a non-vanishing distribution at the position of the emergent operators, there is no condensation of chromoelectric charge there.

\section{Cardy-like limit of the Bethe ansatz equations}
\label{app:Integrability}

This appendix checks that the Cardy-like limit~$\tau\to 0$ of the Bethe ansatz equations of~\cite{Benini:2018mlo} give the same number of solutions as the Bethe roots of the lattice Phase model of~\cite{KORFF2010200}. These are the~$N$ dominating fixed-points~\eqref{SolsCardy} for the limit~$m=1$ and~$n=0\,$. As it will be illustrated below, the generalization of this statement to generic~$m$ and~$n$ is more involved and it will be left for future work. 

Our starting point are the~$SU(N)$ Bethe ansatz equations of~\cite{Benini:2018mlo}
\cite{Benini:2021ano}
\begin{equation}\label{BAEq}
\frac{Q_i}{Q_N}\,-\,1\,=\,0\,,\, \qquad i=1\,\ldots\, N
\,-\,1\,,
\end{equation}
where
\begin{equation}\label{BetheOperator}
Q_i(v,\Delta,\tau)\, \equiv\, e^{-2\pi\i \sum_{j=1}^{N} v_{ij}}\,\prod_{a=1}^3\,\prod_{j=1}^{N}\frac{\theta_0(v_{ji}+\Delta_a,\tau)}{\theta_0(v_{ij}+\Delta_a,\tau)}\,.
\end{equation}
The chemical potentials are assumed to obey the linear constraint
\begin{equation}
\sum_{a=1}^3\Delta_a\,-\,2\tau \= n \in \mathbb{Z}\,.
\end{equation}
Let us define~$z\equiv z(v_{ij})\,=\, v_{ji}\,+\,\Delta_a\,$. For a complex number~$x$ let us define its two real components~$x_{\perp}$ and~$x_{||}$ by the relation~$x=x_{\perp}+x_{||} \widetilde{\tau}$, with~$\widetilde{\tau}\equiv m\tau+n\,$ with~$m>0$ and~$n$ two co-prime integers. 

As explained in the companion paper~\cite{PaperBulk} for generic~$\Delta_{a}$ away from the walls in the complex~$z$-plane at which the component~$\xi_{\ell \perp}$ (with~$\ell \in \mathbb{Z}\,$) of
\be
\xi_{\ell}(z) = z-\ell \tau-\lfloor z_{\perp}- \ell \frac{n}{m}\rfloor -1\,,
\ee
equals~$0$ or~$1\,$,
\begin{equation}\label{IdentityTheta00}
\theta_0(z)\,=\, e^{\pi\text{i}\,\sum_{\ell=0}^{m-1} B_{2,2}(\xi_{\ell}|\widetilde{\tau},-1)\,+\, L{(z)}}
\underset{\tau\to -\frac{n}{m}}{\simeq}\, e^{\pi\,\text{i}\,\sum_{\ell=0}^{m-1} B_{2,2}(\xi_{\ell}|\widetilde{\tau},-1)\,+\,\text{Disc}}\,.
\end{equation}
In this equation
\begin{equation}
B_{2,2}(z-1\,| \,\tau,-1) \,
\equiv\, -\frac{1}{\tau }\, B_2(z)\,+\,B_1(z)\,-\,\frac{\tau }{6}\,.
\end{equation}
Furthermore, the Cardy-like limit of function~$L(z)$ vanishes almost everywhere except for at walls where the limit can happen to be undefined. In the cases where the limit is undefined only the lateral limits to the walls are well-defined and the corresponding discontinuities, denoted as~Disc in~\eqref{IdentityTheta00}, are bound to cancel the discontinuities across the very same walls (branch cuts) that the piecewise polynomial part~$\pi\,\sum_{\ell=0}^{m-1} \text{i}\, B_{2,2}(\xi_{\ell}|\widetilde{\tau},-1)\,$ exhibits.~\footnote{Such cancellation follows from the fact that~$\theta_0(z)$ has no branch cuts in the complex z-plane.} The symbol~$\simeq$ in~\eqref{IdentityTheta00} means equality up to exponentially suppressed corrections in the limit~$\tau\to-\frac{n}{m}$. The objects~$B_1\equiv z-\frac{1}{2}$ and~$B_2 \equiv z^2-z+\frac{1}{6}$ are the first and second Bernoulli polynomials.

From~\eqref{IdentityTheta00} it follows that to keep the~$Q_i$'s, and consequently the~$\frac{Q_i}{Q_N}$ finite and non-vanishing in the limits~$\tau\to-\frac{n}{m}$ one can require
$v_{ij} \,\to\, u_{ij} \widetilde{\tau}\,+\,\mathbb{Z}$ with finite and real~$u_{ij}\,=\,u_i-u_j\,\equiv\, v_{|| ij}\,\neq\,0\,$ for~$i\,\neq\,j\,$. That follows from the fact that in limits~$\tau\to-\frac{n}{m}$ where the real component~$v_{\perp ij}$ takes generic values, the quantity~$\frac{Q_i}{Q_N}$ can not be equal to~$1$ because either it vanishes or diverges.~\footnote{It diverges or vanishes because for generic~$v_{\perp ij}\,\notin\, \mathbb{Z}$ there is a nontrivial component proportional to~$\frac{1}{\widetilde{\tau}}\,$ contributing in the limits~$\tau\to -\frac{n}{m}\,$. This divergent contribution to the exponent vanishes for~$v_{\perp ij}\,\in\,\frac{1}{2}\,+\, \mathbb{Z}$ as well but the contribution to the index coming from the roots that could potentially arise from ansatz obeying~$v_{\perp ij}\,\in\,\frac{1}{2}\,+\, \mathbb{Z}$ can be shown to be exponentially suppressed with respect to the one associated to the ansatz~$v_{\perp ij}\,\in\, \mathbb{Z}$ ones~\eqref{SolsCardy}.} For generic values of~$m$ and~$n$ there are ways to ensure finiteness of~$\frac{Q_i}{Q_N}\,$, for instance by requiring~$v_{ij} \,\to\, u_{ij} \widetilde{\tau}\,+\,\frac{\mathbb{Z}}{m}\,$. As announced, we focus on the case~$m=1$ and~$n=0$ in that case, starting from~\eqref{IdentityTheta00}~\footnote{Working at values of~$\Delta_a$ for which the discontinuities Disc are away from~$v_{\perp}\in\mathbb{Z}\,$.} a computation shows that in the Cardy-like limit the Bethe ansatz equations become
\be\label{BetheAsymptotics}
\frac{Q_i}{Q_N}\= \Bigl(\widetilde{{\zeta_i}}\Bigr)^N\= 1\,, \,\qquad i\=1\,,\,\ldots\,,\,N-1\,,
\ee
where we have defined~$\widetilde{\zeta}_{i}\equiv e^{2 \pi\i\, (u_{i}-u_{N})}\,$. The solutions to~\eqref{BetheAsymptotics} are~$\widetilde{\zeta}_i=$~$N$-roots of unity. Discarding solutions with coincident~$\zeta_i$'s and solutions that are identified after permutations, it follows that there are~$N$ different solutions, which are~\eqref{SolsCardy}.~\footnote{That the solutions to the equations~\eqref{BAEq} in the limit~$\tau\to0$ are related to roots of unity was previously found in~\cite{Hosseini:2016cyf} in a related context.} These are also the Bethe roots of the previously mentioned phase model~\cite{KORFF2010200}\cite{Okuda:2012nx}. For the reasons explained in the main body of the paper we expect the same conclusion to hold for generic Cardy-like limits~$\tau\to -\frac{n}{m}$.

\paragraph{A comment about Bethe/Gauge correspondence} The Bethe roots (or fixed-points) of the system here studied~\cite{Benini:2018mlo}\cite{Cabo-Bizet:2020ewf} are expected to be in one-to-one correspondence with Hamiltonian eigenstates of an integrable system~\cite{Nekrasov:2009uh}. Given the results presented in references~\cite{KORFF2010200,Bogoliubov:1997soj}\cite{Okuda:2012nx}, the proposal just put-forward suggests that in Cardy-like limit such an integrable model could be related to the Phase model studied in~\cite{KORFF2010200} for the values~$z_{\text{there}}=1\,$,~$k_{\text{there}}=1$ particles moving along a circle with~$n_{\text{there}}=N$ sites. This lattice model is solvable by the Algebraic Bethe ansatz method. It was also shown in~\cite{Okuda:2012nx} that the partition function of the~$U(N)_1/U(N)_1$ WZNW model on a Riemann surface of genus~$h$ corresponds to the sum of the Bethe norms (to the power~$1-h$) of the Hamiltonian eigenstates of a Phase model. Thus, on the two-torus ($h=1$) such partition function equals the number of Bethe roots of the algebraic equations~(6.2) of~\cite{KORFF2010200} for~$z_{\text{there}}=1\,$,~$k_{\text{there}}=1\,$, and~$n_{\text{there}}=N$, which again is~$N$. In appendix~\ref{app:Integrability} we check that the limit~$\tau\to 0$ of the Bethe ansatz equations of~\cite{Benini:2018mlo,Benini:2018ywd,Closset:2017bse} gives the same number of solutions as the Bethe roots of the lattice Phase model of~\cite{KORFF2010200} for~$z_{\text{there}}=1\,$,~$k_{\text{there}}=1\,$, and~$n_{\text{there}}=N$. These are the~$N$ dominating fixed-points~\eqref{SolsCardy} for~$m=1$ and~$n=0\,$.~\footnote{We should note that the Bethe ansatz equations (6.2) of~\cite{KORFF2010200} are not the same as the Cardy-like limit of the Bethe ansatz equations of~\cite{Benini:2021ano}, albeit the number of relevant solutions is expected to be the same. The natural guess is that somehow level-rank duality is at play in relating them, as advocated in~\cite{Okuda:2012nx}. We hope to reach a better understanding of this point in the future.}


\providecommand{\href}[2]{#2}\begingroup\raggedright\endgroup

\end{document}